\documentclass[prl, reprint, superscriptaddress, amsmath, amssymb, aps, floatfix, longbibliography, nofootinbib]{revtex4-2}

\usepackage{graphicx}
\usepackage{dcolumn}
\usepackage{bm}
\usepackage{xcolor}
\usepackage{hyperref}
\hypersetup{
  colorlinks=true,
  linkcolor=blue!50!black,
  citecolor=blue!50!black,
  urlcolor=blue!50!black,
}

\begin{document}

\title{Pairing from anti-Hund's coupling under infinite Hubbard repulsion}

\author{Donggyu Kim}
\author{Ya-Hui Zhang}
\affiliation{William H.\ Miller III Department of Physics and Astronomy,
  Johns Hopkins University, Baltimore, Maryland 21218, USA
}

\date{\today}

\begin{abstract}
Electron--phonon coupling is a natural source of superconductivity, but in strongly correlated
materials the phonon-mediated interaction is often much weaker than the on-site Hubbard repulsion $U$.
Whether, and how, it can still produce pairing when $U$ dominates is a basic open question. We address
this question by studying a minimal model with an anti-Hund's coupling $J_A$: an on-site
spin exchange between two orbitals or two valleys, mediated by optical phonons, that favors a
local inter-orbital spin singlet. In the limit $U=+\infty$, the two-orbital Hubbard model
on the triangular lattice, doped away from the insulator at $n=2$ electrons per site, reduces
to a four-flavor $t$--$J_A$ model. Density-matrix renormalization group calculations find robust pairing
throughout $1\le J_A/|t|\le10$, with the largest pairing gap at $J_A\sim|t|$.
The Cooper pairs reside on bonds and do not come from a direct attraction provided by
$J_A$, since infinite $U$ excludes on-site pairs from the Hilbert space. For
$J_A\gg|t|$, the normal state is a second Fermi liquid whose small Fermi surface
violates the perturbative Luttinger theorem, and a strong-coupling expansion in
$t/J_A$  provides a controlled effective theory of superconductivity from this unconventional metal. Our
results provide strong evidence for a strongly coupled superconductor beyond BCS theory,
driven by the interplay of the kinetic energy, the phonon-mediated anti-Hund's
coupling, and the strong Hubbard $U$. Finally, we discuss possible
applications to magic-angle twisted bilayer graphene.
\end{abstract}

\maketitle

\suppressfloats[t]
\textit{Introduction.---}
In conventional superconductors described by the Bardeen--Cooper--Schrieffer (BCS) theory~\cite{BCS1957}, the phonon-mediated attraction overcomes the Coulomb repulsion because of retardation~\cite{MorelAnderson1962}. In strongly correlated materials, however, the on-site Hubbard repulsion $U$ can exceed the phonon-mediated interaction by an order of magnitude or more, and it is far from obvious whether electron--phonon coupling can still produce superconductivity. Phonons that couple to an orbital or valley degree of freedom generate an on-site anti-Hund's coupling $J_A$, which favors a spin singlet of two electrons in different orbitals. On its own, the on-site attraction that $J_A$ provides is still overwhelmed by a large $U$. However, $J_A$ is intrinsically a spin--spin interaction between the two electrons on a doubly occupied site, and this interaction remains fully active however large $U$ is. It is then natural to ask whether this spin interaction can still produce superconductivity under strong Hubbard $U$.

\begin{figure}[t!]
\centering
\includegraphics[width=8.6cm]{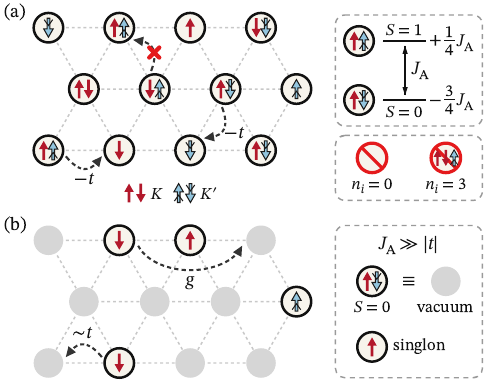}
\caption{Four-flavor $t$--$J_A$ model on the triangular lattice.  (a) Each site holds one electron
(singlon) or two (doublon); an arrow's direction gives the spin and its color indicates the valley, $K$ (red) or
$K'$ (blue, with a chevron). Due to the infinite Hubbard $U$, the $n_i=0$ and $n_i=3$ states (middle right) are excluded from the Hilbert space, so two doped holes cannot share a site.  Dashed arcs are nearest-neighbor hops ($-t$); the crossed arc, a hop onto a doublon,
is thus forbidden.  Top right: on an intervalley
doublon, the anti-Hund's coupling $J_A$ splits the spin triplet ($+J_A/4$) from the spin singlet
($-3J_A/4$).  (b) For $J_A\gg|t|$, each doublon is the intervalley spin singlet of (a), drawn as a gray disk; their product
serves as the vacuum.  Singlons, one per doped hole, form a dilute gas.  Importantly, there is a pivot hopping term $g\sim t^2/J_A$ which rotates a pair of singlons on a bond to another bond. The pivot hopping exists for any spin--valley state of the pair, but provides attraction only for the intervalley spin-singlet pair.}
\label{fig:model}
\end{figure}

The alkali-doped fullerides $A_3$C$_{60}$ are the best-studied example~\cite{Capone2002,Gunnarsson1997RMP,Capone2009RMP}. There, Jahn--Teller vibrations of the C$_{60}$ molecules overcome the Hund's coupling and produce an effective anti-Hund's coupling~\cite{Capone2009RMP,Nomura2015}. Experimentally, $s$-wave superconductivity with $T_c$ up to 38~K appears next to a Mott insulator~\cite{Gunnarsson1997RMP,Ganin2008,Takabayashi2009}. Dynamical mean-field theory (DMFT) studies of multi-orbital Hubbard models with such a coupling found that strong correlations do not suppress this phonon-driven pairing and can even enhance it near the Mott transition~\cite{Capone2002,HanGunnarssonCrespi2003,Capone2009RMP}. A more recent candidate is magic-angle twisted bilayer graphene (TBG), where superconductivity emerges upon doping the correlated insulators at flat-band fillings $\nu=\pm2$~\cite{Cao2018Mott,Cao2018SC,Yankowitz2019,Lu2019}. Its pairing mechanism is a central open question in moir\'e physics. Theories based on phonon-mediated attraction~\cite{WuMacDonaldMartin2018,Lian2019} or weak-coupling Fermi-surface instabilities~\cite{Isobe2018,Kennes2018,Gonzalez2019,YouVishwanath2019} start from a conventional Fermi liquid. Experiments, however, find small Fermi surfaces near the correlated insulators~\cite{Cao2018SC} and a pseudogap that persists when superconductivity is suppressed in TBG~\cite{Oh2021STM} and in twisted trilayer graphene~\cite{Park2026Tunnel}, which point instead to a strongly correlated normal state with a small carrier density. Recent photoemission measurements on superconducting TBG have found evidence that the flat-band electrons couple strongly to an intervalley optical phonon at the graphene $K$ point, while the signature is absent in the non-superconducting samples studied~\cite{Chen2024EPC}. Theoretically, such phonons are predicted to induce an anti-Hund's coupling $J_A$ between the two valleys on each moir\'e site, which favors an intervalley spin singlet and has been proposed as the pairing glue~\cite{BlasonFabrizio2022,WangMolecular2024,WangEPC2025,ZhaoZhang2025}.

The existing DMFT studies~\cite{Capone2002,HanGunnarssonCrespi2003,Capone2009RMP,Nomura2015}, however, treat $U$ of the order of the bandwidth and capture only on-site pairs. In the  limit of infinite  $U$, on-site pairs are excluded altogether, and it is unclear whether $J_A$ can still produce pairing. This limit is relevant to TBG, where the on-site Hubbard repulsion, estimated at $U\approx30$--$60$~meV, far exceeds $J_A\approx1$--$4$~meV~\cite{WangMolecular2024}, and the flat-band kinetic energy may be much smaller than $U$ as well. To establish that a strongly coupled superconductor can exist even at $U=+\infty$, we study a simple model on a topologically trivial band that keeps only the ingredients this question requires: the hopping $t$, the anti-Hund's coupling $J_A$, and an infinite $U$. It thus leaves out the quantum geometry and the competing symmetry-breaking orders of real TBG\@. In return, its small Hilbert space allows unbiased numerical calculations with the density-matrix renormalization group (DMRG) and a controlled analytical theory in the strong-coupling limit.

Concretely, we consider a two-orbital Hubbard model on the triangular lattice and, following TBG, call the two orbitals valleys $K$ and $K'$. Doping the insulator at $n=2$ with $x$ holes per site and taking $U\to\infty$, so that each site holds one or two electrons, yields a four-flavor $t$--$J_A$ model [Fig.~\ref{fig:model}(a)]~\cite{ZhangMao2020,ZhaoZhang2025}. For $J_A\gg|t|$, each doubly occupied site forms an intervalley spin singlet, and the product of these singlets serves as the vacuum. Each doped hole turns one singlet into a \emph{singlon}, a hard-core fermion carrying one of the four spin--valley flavors~\cite{YangOhZhang2024}. The singlons hop with amplitude $\sim t$ and interact on the scale $t^2/J_A$ [Fig.~\ref{fig:model}(b)]. Their Fermi surface encloses only the $x$ doped holes per site rather than the $2-x$ electrons, which defines a second Fermi liquid (sFL)~\cite{ZhangMao2020,WuYangZhang2024,ZhaoZhang2025}.

\begin{figure}[t]
\centering
\includegraphics{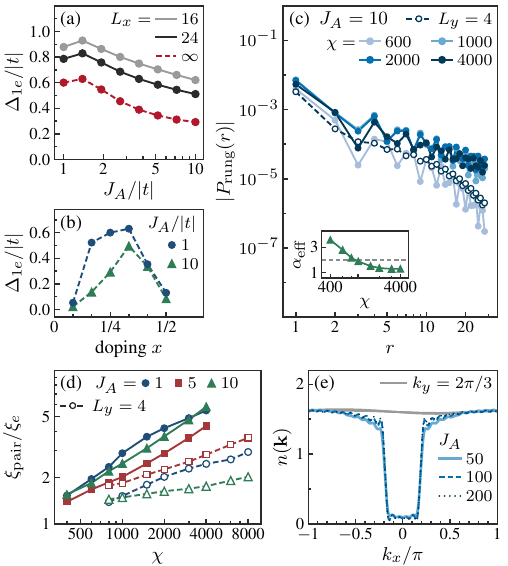}
\caption{\label{fig:2}DMRG results at $x=1/4$, except in (b).  (a) Single-particle gap $\Delta_{1e}$ on $L_y=3$ cylinders at $L_x=16$ and $24$, and its $1/L_x\to0$ extrapolation (dashed), versus $J_A/|t|=10^{k/7}$
($k=0,\ldots,7$).  (b) The $1/L_x\to0$ extrapolation of $\Delta_{1e}$ versus $x$ for $J_A/|t|=1$ and $10$.  (c) Pairing correlation $|P_{\rm rung}(r)|$ from infinite DMRG (iDMRG) at $J_A=10$, with $r$ the distance along the cylinder.  Solid curves: $L_y=3$ at bond dimensions $\chi=600$, $1000$, $2000$, and $4000$; dashed curve with open circles:
$L_y=4$ at $\chi=4000$.  Inset ($L_y=3$): effective exponent $\alpha_{\rm eff}$ versus $\chi$, from a fit $|P_{\rm rung}|\propto r^{-\alpha_{\rm eff}}$ over $8\le r\le24$; the dashed line marks
$\alpha_{\rm eff}=2$.  (d) Ratio $\xi_{\rm pair}/\xi_e$ of the pair to the single-particle correlation length versus $\chi$ for $J_A=1$, $5$, and $10$.  Solid curves with filled symbols: $L_y=3$; dashed curves with open symbols: $L_y=4$.  (e) Momentum distribution $n(\mathbf k)$ from iDMRG on $L_y=3$ at $\chi=4000$ for $J_A=50$, $100$, and $200$, with $k_x$ and $k_y$ the momenta along and around the cylinder.  Blue curves: $k_y=0$ branch; gray curves: $k_y=2\pi/3$ branch.}
\end{figure}

\begin{figure}[t]
\includegraphics[width=\columnwidth]{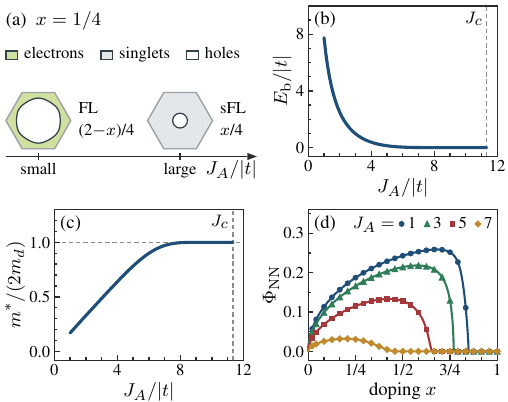}
\caption{\label{fig:mf}(a) Schematic normal-state Fermi surfaces of one flavor at $x=1/4$, areas to scale. In the Fermi liquid (FL), the electrons fill $(2-x)/4$ of the Brillouin zone. In the sFL, the carriers form a hole pocket of area $x/4$ around $\Gamma$. Gray: the on-site singlet vacuum of Fig.~\ref{fig:model}(b). (b) Binding energy $E_{\rm b}$ and (c) effective mass $m^*$ of the two-carrier bound state in the effective model, versus $J_A$. The dashed lines mark $J_c$ and, in (c), $m^*=2m_d$, the mass of two free carriers. (d) Self-consistent mean-field pair amplitude $\Phi_{\rm NN}$ at $T=0$ in the Gutzwiller approximation, versus doping $x$ for $J_A/|t|=1$, $3$, $5$, and $7$.}
\end{figure}

Our DMRG calculations~\cite{White1992} indicate strong pairing throughout $1\le J_A/|t|\le10$, with the pairing gap largest at $J_A\sim|t|$ [Fig.~\ref{fig:2}(a)]. Since the gap does not keep growing with $J_A$, the pairing is not a direct attraction from $J_A$. Instead, a pair resides on a bond, and its formation necessarily involves the hopping $t$. In the strong-coupling expansion for $J_A\gg|t|$, we derive a controlled effective theory and solve it exactly for two carriers, which bind into an intervalley spin-singlet pair with extended-$s$ symmetry for $J_A<J_c\approx11.3|t|$. Both DMRG and a mean-field treatment of this theory find that the pairing grows with $x$ at small hole density [Figs.~\ref{fig:2}(b) and \ref{fig:mf}(d)], similar to the superconducting gap reported experimentally in magic-angle twisted  trilayer graphene~\cite{Park2026Tunnel}. Our results therefore provide strong evidence that $J_A$ and $t$ alone can produce superconductivity at infinite $U$, with pairs that reside on bonds, unlike the on-site pairs of DMFT at moderate $U$~\cite{Capone2002,HanGunnarssonCrespi2003,Capone2009RMP}. We also suggest a possible transition from a small to a large Fermi surface in the normal state near $J_A\sim|t|$, where the pairing is strongest. Finally, we discuss possible applications to TBG, where the effective hopping arises from a combination of interaction and quantum geometry.

\textit{Model.---} We start from a two-orbital Hubbard model on the triangular lattice with an on-site anti-Hund's coupling $J_A$. In the limit $U\to\infty$, it reduces to the four-flavor $t$--$J_A$ model~\cite{SM},
\begin{equation}
H=-t\sum_{\langle ij\rangle,\alpha}\left(\tilde c^{\dagger}_{i\alpha}\tilde c_{j\alpha}
+\mathrm{H.c.}\right)+J_A\sum_i \mathbf S_{iK}\cdot\mathbf S_{iK'} ,
\label{eq:H}
\end{equation}
where the flavor index $\alpha=(\tau,\sigma)$ combines the valley $\tau=K,K'$ and the spin
$\sigma=\uparrow,\downarrow$, and $\langle ij\rangle$ runs over nearest-neighbor bonds. Here
$\tilde c_{i\alpha}=P_i c_{i\alpha}P_i$, where $P_i$ restricts the site occupation to
$n_i\in\{1,2\}$. The spin
operator of valley $\tau$ is $\mathbf S_{i\tau}=\tfrac12\sum_{\sigma\sigma'}
c^{\dagger}_{i\tau\sigma}\bm\sigma_{\sigma\sigma'}c_{i\tau\sigma'}$ with $\bm\sigma$ the Pauli
matrices. We take $J_A>0$ throughout. Since the triangular lattice is not bipartite, the
sign of $t$ is physical. In this work we take $t=-1$ and quote all energies in units of $|t|$.

The projection leaves ten states per site: four singlons ($n_i=1$) and six doublons ($n_i=2$) [Fig.~\ref{fig:model}(a)].
The doublons consist of one intervalley spin singlet, three intervalley spin triplets, and two
intravalley singlets. The coupling $J_A$ acts only when both valleys are occupied: the on-site energy is $-3J_A/4$
for the intervalley singlet, $+J_A/4$ for the intervalley triplet, and $0$ for singlons and
intravalley doublons. An electron hops only from a doublon onto a
neighboring singlon, keeping its flavor. The model conserves the valley populations $N_K$ and $N_{K'}$ separately and is invariant under global
spin rotations, so $S^z$ is conserved as well.

We work at density $n=2-x$ per site, so that $x\in[0,1]$ is the fraction of singlons. The projected Hamiltonian contains no $U$, so the
model has only two parameters, $J_A/|t|$ and the doping $x$. Unlike the superexchange $4t^2/U$
of the standard $t$--$J$ model, which vanishes as $U\to\infty$, $J_A$ is an on-site spin coupling from optical phonons and thus can in principle be larger than $|t|$ in flat-band systems.

\textit{DMRG results.---} We first demonstrate pairing in this model using DMRG~\cite{White1992,White1993} with the TeNPy package~\cite{Hauschild2018,Hauschild2024}.
To characterize the pairing strength, we compute the single-particle gap $\Delta_{1e}=E_0(N{+}1)+E_0(N{-}1)-2E_0(N)$ at $x=1/4$,
with $E_0(N)$ the ground-state energy of $N$ electrons.  We use finite DMRG on $L_y=3$ cylinders at bond dimension
$\chi=1500$.  The gap is computed at $L_x=16$ and $24$ and extrapolated in $1/L_x$ [Fig.~\ref{fig:2}(a)].
$\Delta_{1e}$ is finite throughout $1\le J_A/|t|\le10$ and is largest at $J_A\approx1.4|t|$.  To study the doping dependence, we show $\Delta_{1e}$ for $1/12\le x\le1/2$ at $J_A/|t|=1$ and $10$ in Fig.~\ref{fig:2}(b).  At small $x$, $\Delta_{1e}$ increases with doping for both couplings.  The gap is largest at $x\approx1/4$--$1/3$ and decreases at larger $x$.  Such a dome of the gap resembles the superconducting gap reported in spectroscopic experiments on magic-angle twisted trilayer graphene~\cite{Park2026Tunnel}.  Note that it is quite striking to have a pairing gap increasing with hole doping at small $x$. For example, in the hole-doped cuprates, the pairing gap usually decreases with the hole doping $x$ in the underdoped regime~\cite{Miyakawa1998,LeeNagaosaWen2006}, although $T_c$ increases with $x$ because it is limited by the phase stiffness.  In this sense, the dome of the pairing gap in twisted trilayer graphene is quite unusual and not expected in most theories. A pairing gap that increases with doping at small $x$ was also found in the empty--singlon--doublon (ESD) model for the kinetic superconductor~\cite{OhYangZhangKineticSC} and in a parton mean-field theory of TBG~\cite{ZhaoZhang2025}. Later we will show that the kinetic energy is also essential for our pairing mechanism.

To probe the pairing correlations directly, we further perform infinite DMRG (iDMRG) calculations~\cite{McCulloch2008} on $L_y=3$ and $4$ cylinders at $x=1/4$ and bond dimensions up to $\chi=4000$ and $8000$, respectively. We measure the pairing correlation $P_{\rm rung}(r)=\langle\Delta^\dagger(0)\,\Delta(r)\rangle$, where the pairing operator
\begin{equation}
\Delta^\dagger(i_x)=\tilde c^{\dagger}_{(i_x,0)K\uparrow}\tilde c^{\dagger}_{(i_x,1)K'\downarrow}
-\tilde c^{\dagger}_{(i_x,0)K\downarrow}\tilde c^{\dagger}_{(i_x,1)K'\uparrow}
\label{eq:pairop}
\end{equation}
creates an intervalley spin-singlet pair on the nearest-neighbor bond between the sites $(i_x,0)$ and $(i_x,1)$, labeled by their coordinates along and around the cylinder, and $r$ is the distance between the two bonds along the cylinder.  Figure~\ref{fig:2}(c) shows that the pairing correlation at $J_A=10$ exhibits power-law behavior with increasing bond dimension $\chi$.  A power-law fit on $L_y=3$ at $\chi=4000$ gives an exponent $\alpha_{\rm eff}\simeq1.2$--$1.3$ [inset of Fig.~\ref{fig:2}(c)], well below $2$, so the pair susceptibility diverges at low temperature~\cite{JiangWengKivelson2018}.  We also find that the ratio $\xi_{\rm pair}/\xi_e$ of the pair correlation length to the single-particle correlation length increases with $\chi$ for $J_A=1$, $5$, and $10$ on both $L_y=3$ and $4$ [Fig.~\ref{fig:2}(d)], although $\xi_{\rm pair}$ is still short for $J_A<10$.  All of these results support the existence of pairing in this model. From Figs.~\ref{fig:2}(a) and \ref{fig:2}(b), $\Delta_{1e}$ reaches about $0.6|t|$, implying a huge pairing gap of order $|t|$.

\textit{Analytical theory at large $J_A$.---} We now provide an analytical theory of the pairing mechanism in the strong-coupling limit.
An expansion in $t/J_A$ shows how the hopping makes two singlons attract.  At $x=0$, every site is doubly occupied and hopping is forbidden due to the infinite Hubbard $U$, so the ground state is exactly a product state of on-site singlets for any $J_A>0$.  Upon hole doping, we refer to the singlons on the singlet vacuum of Fig.~\ref{fig:model}(b) as carriers.  At first order, a carrier hops with amplitude $t/2$ by trading places with a neighboring singlet, keeping its flavor.  The $x$ carriers per site fill four degenerate bands and form the small Fermi surface of the sFL [Fig.~\ref{fig:mf}(a)]~\cite{WuYangZhang2024,ZhaoZhang2025}.  Indeed, at $J_A=50$, $100$, and $200$, far above $J_c$, the momentum distribution $n(\mathbf k)$ from iDMRG is depleted only in a small hole pocket around $\Gamma$ [Fig.~\ref{fig:2}(e)], with the size expected if the Fermi surface encloses only the carriers~\cite{SM}.

At second order in $t/J_A$~\cite{ChaoSpalekOles1977,MacDonaldGirvinYoshioka1988,CrepelFu2021}, there are two additional terms, and we obtain the following effective Hamiltonian~\cite{SM}:
\begin{equation}
\begin{aligned}
\hat H_{\rm eff}={}&\frac t2\sum_{\langle ij\rangle\beta}\bigl(d^\dagger_{i\beta}d_{j\beta}+{\rm H.c.}\bigr)+V_{\rm NN}\sum_{\langle ij\rangle}n^d_in^d_j\\
&+\sum_w g_w\sum_p\sum_{\substack{q,j\,\sim\,p\\ j\neq q}}B^\dagger_{w,pj}B^{\phantom\dagger}_{w,pq}.
\end{aligned}
\label{eq:heffmain}
\end{equation}
Here $d^\dagger_{i\beta}$ creates a carrier with flavor $\beta$ on site $i$, and $n^d_i$ is the number of carriers on site $i$. We have the hard-core constraint $n^d_i\le1$ at each site $i$, because two carriers on the same site would correspond to the excluded empty state $n_i=0$ [Fig.~\ref{fig:model}(a)]. The Hilbert space at each site is thus $1+4=5$ dimensional: the singlet vacuum and a carrier with one of the four flavors. The first term represents the kinetic energy of the carriers. The second term represents a nearest-neighbor repulsion between the carriers, with $V_{\rm NN}=\frac{17}{6}t^2/J_A$. The third term is the pivot hopping term. The operator $B^\dagger_{w,pq}$ creates a pair of carriers on the bond $pq$ in the spin--valley state $w$, so this term rotates the pair from the bond $pq$ to the bond $pj$ around the site $p$ [Fig.~\ref{fig:model}(b)], where $q$ and $j$ are two different neighbors of $p$. Note that $g_w$ is negative only for the valley-symmetric intervalley spin-singlet pair, for which $g_w=-\frac{17}{12}t^2/J_A$~\cite{SM}. As we will show, the pivot hopping term overcomes the repulsion $V_{\rm NN}$ and provides a net attraction for this pair.

\textit{Exact solution for two singlons.---} We first solve the resulting effective model exactly for two singlons~\cite{SM}.  In relative coordinates and at zero total momentum, the two-carrier problem is mapped to a single-particle hopping problem on the triangular lattice [see Fig.~\ref{fig:oneparticle} of the Supplemental Material]. The particle hops with amplitude $t$, and the origin is removed due to the hard-core constraint. The pivot hopping term gives an additional hopping between any two of the six sites neighboring the origin, which also have the potential energy $V_{\rm NN}$. For this pair, the additional hopping has the same sign as $t$. The hopping is thus stronger around the origin, and the particle tends to localize there, corresponding to a bound pair. We find that the ground state is a bound state only for this pair, with extended-$s$ symmetry, and only for $J_A<J_c=\frac{34}{3}|t|\approx11.3|t|$.  We emphasize that the attraction is generated by the hopping $t$: at a fixed $J_A$, two carriers bind only when the hopping is large enough. In this sense, the kinetic energy is essential for our pairing mechanism.  The bound pair survives an additional nearest-neighbor repulsion $V$ up to $V_c=\frac{34}{3}t^2/J_A-|t|$~\cite{SM}.

The binding energy $E_{\rm b}$ grows as $J_A$ decreases [Fig.~\ref{fig:mf}(b)].  This is qualitatively consistent with our DMRG results, where the pairing gap decreases with $J_A$ for $J_A\gtrsim1.4|t|$ [Fig.~\ref{fig:2}(a)].  Note that the expansion holds for $J_A\gg|t|$, so the values at small $J_A$ are only qualitative.  A tightly bound pair is about one bond long~\cite{SM}.  As $J_A\to J_c$, $E_{\rm b}$ becomes exponentially small and the pair grows rapidly.

The bound pair is lighter than two free carriers, whose mass $2m_d$ does not depend on $J_A$.  Its effective mass $m^*$ stays below $2m_d$ for all $J_A<J_c$ and approaches $2m_d$ at $J_c$ [Fig.~\ref{fig:mf}(c)].  The pair is light because the pivot hop that binds it also moves it.  A smaller $J_A$ strengthens this hop and makes the pair both more tightly bound and lighter.  This is contrary to the naive expectation that a more tightly bound pair is heavier, as in the attractive Hubbard model~\cite{ChienChenLevin2008,SM}.

\textit{Renormalized mean-field theory.---} The two-carrier solution above describes the limit $x\to0$. We now move to the case with a finite density of carriers. The problem is then not exactly solvable, and we treat the effective model within a renormalized mean-field theory, with the hard-core constraint handled in the Gutzwiller approximation~\cite{ZhangGrosRiceShiba1988,SM}.  The Gutzwiller factors renormalize the hopping and the pivot hopping terms. After a mean-field decoupling in the intervalley spin-singlet channel, we obtain the following mean-field Hamiltonian:
\begin{equation}
\begin{aligned}
\hat H_{\rm MF}={}&\sum_{\mathbf k\beta}\xi_{\mathbf k}\,d^\dagger_{\mathbf k\beta}d_{\mathbf k\beta}\\
&+\frac12\sum_{\mathbf k\beta\gamma}\Bigl(D(\mathbf k)\,[w]_{\beta\gamma}\,d^\dagger_{\mathbf k\beta}d^\dagger_{-\mathbf k\gamma}+{\rm H.c.}\Bigr).
\end{aligned}
\label{eq:HMFmain}
\end{equation}
Here $\xi_{\mathbf k}$ is the renormalized dispersion and $[w]_{\beta\gamma}$ is the flavor wave function of the valley-symmetric intervalley spin-singlet pair. The gap function is $D(\mathbf k)=-\lambda\,\Phi_{\rm NN}\,[\gamma_a(\mathbf k)-c]$, where the nearest-neighbor pair amplitude $\Phi_{\rm NN}$ is determined self-consistently and $\gamma_a(\mathbf k)=2\sum_{\ell=1}^{3}\cos(\mathbf k\cdot\mathbf e_\ell)$, with $\mathbf e_\ell$ the three nearest-neighbor vectors. The constant $c$ is fixed by requiring that the on-site pair amplitude vanish, since two carriers cannot share a site and the Gutzwiller approximation imposes this constraint only on average~\cite{SM}. The effective coupling is $\lambda=\frac{17}{6}(1-5\kappa_P)\,t^2/J_A$, where $\kappa_P=16(1-x)/(4-x)^2$ is the Gutzwiller factor of the pivot hopping term. The coupling is attractive ($\lambda<0$) for $x<0.88$~\cite{SM}.  In the dilute limit $x\to0$, the gap equation reduces to the two-carrier problem of the previous section, and we recover the exact $E_{\rm b}$ and $J_c$. At finite $x$, pairing occurs only for $J_A<J_c(x)$, where the threshold $J_c(x)$ decreases with doping from $\frac{34}{3}|t|$ at $x\to0$ to $9.5|t|$ at $x=1/4$~\cite{SM}. We note that the mean-field theory underestimates the threshold at finite $x$: DMRG finds pairing at $J_A=10|t|$ and $x=1/4$ [Fig.~\ref{fig:2}], above $J_c(1/4)$. The pair amplitude $\Phi_{\rm NN}$ is finite over a wide range of doping and decreases as $J_A$ grows [Fig.~\ref{fig:mf}(d)].

We now offer intuitions on why the pairing gap increases with $x$ at small doping.  In the hole-doped cuprates, the pairing originates from the spin-singlet correlation of the parent Mott insulator~\cite{LeeNagaosaWen2006} and competes with the kinetic energy of the doped holes, so the pairing gap decreases with $x$.  In contrast, in our model there are only immobile on-site singlets at $x=0$, and the pairs are formed by the doped carriers through the pivot hopping term.  Since this term comes from the hopping $t$, the kinetic energy assists the pairing instead of competing with it.  The pivot hopping acts only on a pair of carriers on neighboring sites.  With more carriers, there are more such pairs, and the pairing gap increases.  In our mean-field theory, all carriers are bound into pairs in the dilute limit, so their density $x$ is proportional to $\Phi_{\rm NN}^2$ and $\Phi_{\rm NN}\propto\sqrt{x}$ [Fig.~\ref{fig:mf}(d)]~\cite{SM}.  As a result, the gap function $D(\mathbf k)\propto\Phi_{\rm NN}$ grows with $x$, and the mean-field single-particle gap increases from the two-carrier binding energy $E_{\rm b}$ at $x\to0$~\cite{SM}.  At larger $x$, the pivot hopping is suppressed because fewer singlets are left on the neighboring sites, and both $\Phi_{\rm NN}$ and the gap show a dome, in qualitative agreement with our DMRG results [Fig.~\ref{fig:2}(b)].

\textit{Discussion.---}
In summary, we have demonstrated robust pairing in the four-flavor $t$--$J_A$ model even at $U=+\infty$.  Interestingly, the pairing does not come directly from the attraction provided by $J_A$. Instead, the Cooper pair resides on a bond, and the kinetic term plays an essential role in the formation of the pair, not just its mobility. In the large-$J_A$ limit, we provide an analytical theory of superconductivity out of the small Fermi surface in the second Fermi liquid (sFL) phase~\cite{ZhangMao2020,YangOhZhang2024}. In contrast, a conventional Fermi liquid (FL) is expected at small $J_A$, so a transition from a small to a large Fermi surface should occur as $J_A$ decreases.  Since the pairing gap is largest at $J_A\sim|t|$, we conjecture that the transition lies nearby. The two Fermi liquids can be distinguished only above the pairing gap, so we leave this transition to a finite-temperature study.

Our theory may have important implications for superconductivity in TBG. It has already been proposed that the superconductor emerges from the sFL phase close to its transition to the FL~\cite{ZhaoZhang2025}, but an unbiased calculation is absent. Our concrete model realizes this scenario in its simplest form: the pairing is robust even with infinite $U$, develops out of a small Fermi surface, and becomes stronger as $t/J_A$ increases, up to $J_A\sim|t|$.  In TBG, the bare band is almost flat, but the interaction can generate a dispersive lower Hubbard band because of the concentrated quantum geometry~\cite{Ledwith2025,ZhaoZhouZhang2025}, which may provide an effective hopping $t_{\mathrm{eff}}$. Therefore our current $t$--$J_A$ model may already capture the essence of the superconductivity in TBG. If we include two orbitals in addition to spin and valley, it is straightforward to generalize our analytical theory here to describe a nematic nodal superconductor with intra-orbital intervalley spin-singlet pairing~\cite{ZhaoZhang2025,WangMolecular2024}. We hope to study the effect of quantum geometry and the competition with flavor-symmetry-breaking phases such as the incommensurate Kekul\'e spiral order~\cite{Kwan2021,Nuckolls2023,Kim2023IVC} using a more realistic model for TBG in the near future.

\textit{Acknowledgments.---} This work was supported by the National Science Foundation under Grant No.~DMR-2237031.

\bibliography{refs}

\begin{thebibliography}{49}%
\makeatletter
\providecommand \@ifxundefined [1]{%
 \@ifx{#1\undefined}
}%
\providecommand \@ifnum [1]{%
 \ifnum #1\expandafter \@firstoftwo
 \else \expandafter \@secondoftwo
 \fi
}%
\providecommand \@ifx [1]{%
 \ifx #1\expandafter \@firstoftwo
 \else \expandafter \@secondoftwo
 \fi
}%
\providecommand \natexlab [1]{#1}%
\providecommand \enquote  [1]{``#1''}%
\providecommand \bibnamefont  [1]{#1}%
\providecommand \bibfnamefont [1]{#1}%
\providecommand \citenamefont [1]{#1}%
\providecommand \href@noop [0]{\@secondoftwo}%
\providecommand \href [0]{\begingroup \@sanitize@url \@href}%
\providecommand \@href[1]{\@@startlink{#1}\@@href}%
\providecommand \@@href[1]{\endgroup#1\@@endlink}%
\providecommand \@sanitize@url [0]{\catcode `\\12\catcode `\$12\catcode `\&12\catcode `\#12\catcode `\^12\catcode `\_12\catcode `\%12\relax}%
\providecommand \@@startlink[1]{}%
\providecommand \@@endlink[0]{}%
\providecommand \url  [0]{\begingroup\@sanitize@url \@url }%
\providecommand \@url [1]{\endgroup\@href {#1}{\urlprefix }}%
\providecommand \urlprefix  [0]{URL }%
\providecommand \Eprint [0]{\href }%
\providecommand \doibase [0]{https://doi.org/}%
\providecommand \selectlanguage [0]{\@gobble}%
\providecommand \bibinfo  [0]{\@secondoftwo}%
\providecommand \bibfield  [0]{\@secondoftwo}%
\providecommand \translation [1]{[#1]}%
\providecommand \BibitemOpen [0]{}%
\providecommand \bibitemStop [0]{}%
\providecommand \bibitemNoStop [0]{.\EOS\space}%
\providecommand \EOS [0]{\spacefactor3000\relax}%
\providecommand \BibitemShut  [1]{\csname bibitem#1\endcsname}%
\let\auto@bib@innerbib\@empty
\bibitem [{\citenamefont {Bardeen}\ \emph {et~al.}(1957)\citenamefont {Bardeen}, \citenamefont {Cooper},\ and\ \citenamefont {Schrieffer}}]{BCS1957}%
  \BibitemOpen
  \bibfield  {author} {\bibinfo {author} {\bibfnamefont {J.}~\bibnamefont {Bardeen}}, \bibinfo {author} {\bibfnamefont {L.~N.}\ \bibnamefont {Cooper}},\ and\ \bibinfo {author} {\bibfnamefont {J.~R.}\ \bibnamefont {Schrieffer}},\ }\bibfield  {title} {\bibinfo {title} {Theory of superconductivity},\ }\href {https://doi.org/10.1103/PhysRev.108.1175} {\bibfield  {journal} {\bibinfo  {journal} {Phys. Rev.}\ }\textbf {\bibinfo {volume} {108}},\ \bibinfo {pages} {1175} (\bibinfo {year} {1957})}\BibitemShut {NoStop}%
\bibitem [{\citenamefont {Morel}\ and\ \citenamefont {Anderson}(1962)}]{MorelAnderson1962}%
  \BibitemOpen
  \bibfield  {author} {\bibinfo {author} {\bibfnamefont {P.}~\bibnamefont {Morel}}\ and\ \bibinfo {author} {\bibfnamefont {P.~W.}\ \bibnamefont {Anderson}},\ }\bibfield  {title} {\bibinfo {title} {Calculation of the superconducting state parameters with retarded electron-phonon interaction},\ }\href {https://doi.org/10.1103/PhysRev.125.1263} {\bibfield  {journal} {\bibinfo  {journal} {Phys. Rev.}\ }\textbf {\bibinfo {volume} {125}},\ \bibinfo {pages} {1263} (\bibinfo {year} {1962})}\BibitemShut {NoStop}%
\bibitem [{\citenamefont {Capone}\ \emph {et~al.}(2002)\citenamefont {Capone}, \citenamefont {Fabrizio}, \citenamefont {Castellani},\ and\ \citenamefont {Tosatti}}]{Capone2002}%
  \BibitemOpen
  \bibfield  {author} {\bibinfo {author} {\bibfnamefont {M.}~\bibnamefont {Capone}}, \bibinfo {author} {\bibfnamefont {M.}~\bibnamefont {Fabrizio}}, \bibinfo {author} {\bibfnamefont {C.}~\bibnamefont {Castellani}},\ and\ \bibinfo {author} {\bibfnamefont {E.}~\bibnamefont {Tosatti}},\ }\bibfield  {title} {\bibinfo {title} {Strongly correlated superconductivity},\ }\href {https://doi.org/10.1126/science.1071122} {\bibfield  {journal} {\bibinfo  {journal} {Science}\ }\textbf {\bibinfo {volume} {296}},\ \bibinfo {pages} {2364} (\bibinfo {year} {2002})}\BibitemShut {NoStop}%
\bibitem [{\citenamefont {Gunnarsson}(1997)}]{Gunnarsson1997RMP}%
  \BibitemOpen
  \bibfield  {author} {\bibinfo {author} {\bibfnamefont {O.}~\bibnamefont {Gunnarsson}},\ }\bibfield  {title} {\bibinfo {title} {Superconductivity in fullerides},\ }\href {https://doi.org/10.1103/RevModPhys.69.575} {\bibfield  {journal} {\bibinfo  {journal} {Rev. Mod. Phys.}\ }\textbf {\bibinfo {volume} {69}},\ \bibinfo {pages} {575} (\bibinfo {year} {1997})}\BibitemShut {NoStop}%
\bibitem [{\citenamefont {Capone}\ \emph {et~al.}(2009)\citenamefont {Capone}, \citenamefont {Fabrizio}, \citenamefont {Castellani},\ and\ \citenamefont {Tosatti}}]{Capone2009RMP}%
  \BibitemOpen
  \bibfield  {author} {\bibinfo {author} {\bibfnamefont {M.}~\bibnamefont {Capone}}, \bibinfo {author} {\bibfnamefont {M.}~\bibnamefont {Fabrizio}}, \bibinfo {author} {\bibfnamefont {C.}~\bibnamefont {Castellani}},\ and\ \bibinfo {author} {\bibfnamefont {E.}~\bibnamefont {Tosatti}},\ }\bibfield  {title} {\bibinfo {title} {Colloquium: {M}odeling the unconventional superconducting properties of expanded {$A_3$C$_{60}$} fullerides},\ }\href {https://doi.org/10.1103/RevModPhys.81.943} {\bibfield  {journal} {\bibinfo  {journal} {Rev. Mod. Phys.}\ }\textbf {\bibinfo {volume} {81}},\ \bibinfo {pages} {943} (\bibinfo {year} {2009})}\BibitemShut {NoStop}%
\bibitem [{\citenamefont {Nomura}\ \emph {et~al.}(2015)\citenamefont {Nomura}, \citenamefont {Sakai}, \citenamefont {Capone},\ and\ \citenamefont {Arita}}]{Nomura2015}%
  \BibitemOpen
  \bibfield  {author} {\bibinfo {author} {\bibfnamefont {Y.}~\bibnamefont {Nomura}}, \bibinfo {author} {\bibfnamefont {S.}~\bibnamefont {Sakai}}, \bibinfo {author} {\bibfnamefont {M.}~\bibnamefont {Capone}},\ and\ \bibinfo {author} {\bibfnamefont {R.}~\bibnamefont {Arita}},\ }\bibfield  {title} {\bibinfo {title} {Unified understanding of superconductivity and {Mott} transition in alkali-doped fullerides from first principles},\ }\href {https://doi.org/10.1126/sciadv.1500568} {\bibfield  {journal} {\bibinfo  {journal} {Sci. Adv.}\ }\textbf {\bibinfo {volume} {1}},\ \bibinfo {pages} {e1500568} (\bibinfo {year} {2015})}\BibitemShut {NoStop}%
\bibitem [{\citenamefont {Ganin}\ \emph {et~al.}(2008)\citenamefont {Ganin}, \citenamefont {Takabayashi}, \citenamefont {Khimyak}, \citenamefont {Margadonna}, \citenamefont {Tamai}, \citenamefont {Rosseinsky},\ and\ \citenamefont {Prassides}}]{Ganin2008}%
  \BibitemOpen
  \bibfield  {author} {\bibinfo {author} {\bibfnamefont {A.~Y.}\ \bibnamefont {Ganin}}, \bibinfo {author} {\bibfnamefont {Y.}~\bibnamefont {Takabayashi}}, \bibinfo {author} {\bibfnamefont {Y.~Z.}\ \bibnamefont {Khimyak}}, \bibinfo {author} {\bibfnamefont {S.}~\bibnamefont {Margadonna}}, \bibinfo {author} {\bibfnamefont {A.}~\bibnamefont {Tamai}}, \bibinfo {author} {\bibfnamefont {M.~J.}\ \bibnamefont {Rosseinsky}},\ and\ \bibinfo {author} {\bibfnamefont {K.}~\bibnamefont {Prassides}},\ }\bibfield  {title} {\bibinfo {title} {Bulk superconductivity at 38 {K} in a molecular system},\ }\href {https://doi.org/10.1038/nmat2179} {\bibfield  {journal} {\bibinfo  {journal} {Nat. Mater.}\ }\textbf {\bibinfo {volume} {7}},\ \bibinfo {pages} {367} (\bibinfo {year} {2008})}\BibitemShut {NoStop}%
\bibitem [{\citenamefont {Takabayashi}\ \emph {et~al.}(2009)\citenamefont {Takabayashi}, \citenamefont {Ganin}, \citenamefont {Jegli{\v c}}, \citenamefont {Ar{\v c}on}, \citenamefont {Takano}, \citenamefont {Iwasa}, \citenamefont {Ohishi}, \citenamefont {Takata}, \citenamefont {Takeshita}, \citenamefont {Prassides},\ and\ \citenamefont {Rosseinsky}}]{Takabayashi2009}%
  \BibitemOpen
  \bibfield  {author} {\bibinfo {author} {\bibfnamefont {Y.}~\bibnamefont {Takabayashi}}, \bibinfo {author} {\bibfnamefont {A.~Y.}\ \bibnamefont {Ganin}}, \bibinfo {author} {\bibfnamefont {P.}~\bibnamefont {Jegli{\v c}}}, \bibinfo {author} {\bibfnamefont {D.}~\bibnamefont {Ar{\v c}on}}, \bibinfo {author} {\bibfnamefont {T.}~\bibnamefont {Takano}}, \bibinfo {author} {\bibfnamefont {Y.}~\bibnamefont {Iwasa}}, \bibinfo {author} {\bibfnamefont {Y.}~\bibnamefont {Ohishi}}, \bibinfo {author} {\bibfnamefont {M.}~\bibnamefont {Takata}}, \bibinfo {author} {\bibfnamefont {N.}~\bibnamefont {Takeshita}}, \bibinfo {author} {\bibfnamefont {K.}~\bibnamefont {Prassides}},\ and\ \bibinfo {author} {\bibfnamefont {M.~J.}\ \bibnamefont {Rosseinsky}},\ }\bibfield  {title} {\bibinfo {title} {The disorder-free non-{BCS} superconductor {Cs$_3$C$_{60}$} emerges from an antiferromagnetic insulator parent state},\ }\href {https://doi.org/10.1126/science.1169163} {\bibfield  {journal} {\bibinfo  {journal} {Science}\ }\textbf {\bibinfo {volume} {323}},\ \bibinfo {pages} {1585} (\bibinfo {year} {2009})}\BibitemShut {NoStop}%
\bibitem [{\citenamefont {Han}\ \emph {et~al.}(2003)\citenamefont {Han}, \citenamefont {Gunnarsson},\ and\ \citenamefont {Crespi}}]{HanGunnarssonCrespi2003}%
  \BibitemOpen
  \bibfield  {author} {\bibinfo {author} {\bibfnamefont {J.~E.}\ \bibnamefont {Han}}, \bibinfo {author} {\bibfnamefont {O.}~\bibnamefont {Gunnarsson}},\ and\ \bibinfo {author} {\bibfnamefont {V.~H.}\ \bibnamefont {Crespi}},\ }\bibfield  {title} {\bibinfo {title} {Strong superconductivity with local {Jahn-Teller} phonons in {C$_{60}$} solids},\ }\href {https://doi.org/10.1103/PhysRevLett.90.167006} {\bibfield  {journal} {\bibinfo  {journal} {Phys. Rev. Lett.}\ }\textbf {\bibinfo {volume} {90}},\ \bibinfo {pages} {167006} (\bibinfo {year} {2003})}\BibitemShut {NoStop}%
\bibitem [{\citenamefont {Cao}\ \emph {et~al.}(2018{\natexlab{a}})\citenamefont {Cao}, \citenamefont {Fatemi}, \citenamefont {Demir}, \citenamefont {Fang}, \citenamefont {Tomarken}, \citenamefont {Luo}, \citenamefont {Sanchez-Yamagishi}, \citenamefont {Watanabe}, \citenamefont {Taniguchi}, \citenamefont {Kaxiras}, \citenamefont {Ashoori},\ and\ \citenamefont {Jarillo-Herrero}}]{Cao2018Mott}%
  \BibitemOpen
  \bibfield  {author} {\bibinfo {author} {\bibfnamefont {Y.}~\bibnamefont {Cao}}, \bibinfo {author} {\bibfnamefont {V.}~\bibnamefont {Fatemi}}, \bibinfo {author} {\bibfnamefont {A.}~\bibnamefont {Demir}}, \bibinfo {author} {\bibfnamefont {S.}~\bibnamefont {Fang}}, \bibinfo {author} {\bibfnamefont {S.~L.}\ \bibnamefont {Tomarken}}, \bibinfo {author} {\bibfnamefont {J.~Y.}\ \bibnamefont {Luo}}, \bibinfo {author} {\bibfnamefont {J.~D.}\ \bibnamefont {Sanchez-Yamagishi}}, \bibinfo {author} {\bibfnamefont {K.}~\bibnamefont {Watanabe}}, \bibinfo {author} {\bibfnamefont {T.}~\bibnamefont {Taniguchi}}, \bibinfo {author} {\bibfnamefont {E.}~\bibnamefont {Kaxiras}}, \bibinfo {author} {\bibfnamefont {R.~C.}\ \bibnamefont {Ashoori}},\ and\ \bibinfo {author} {\bibfnamefont {P.}~\bibnamefont {Jarillo-Herrero}},\ }\bibfield  {title} {\bibinfo {title} {Correlated insulator behaviour at half-filling in magic-angle graphene superlattices},\ }\href {https://doi.org/10.1038/nature26154} {\bibfield  {journal} {\bibinfo  {journal} {Nature}\ }\textbf {\bibinfo {volume} {556}},\ \bibinfo {pages} {80} (\bibinfo {year} {2018}{\natexlab{a}})}\BibitemShut {NoStop}%
\bibitem [{\citenamefont {Cao}\ \emph {et~al.}(2018{\natexlab{b}})\citenamefont {Cao}, \citenamefont {Fatemi}, \citenamefont {Fang}, \citenamefont {Watanabe}, \citenamefont {Taniguchi}, \citenamefont {Kaxiras},\ and\ \citenamefont {Jarillo-Herrero}}]{Cao2018SC}%
  \BibitemOpen
  \bibfield  {author} {\bibinfo {author} {\bibfnamefont {Y.}~\bibnamefont {Cao}}, \bibinfo {author} {\bibfnamefont {V.}~\bibnamefont {Fatemi}}, \bibinfo {author} {\bibfnamefont {S.}~\bibnamefont {Fang}}, \bibinfo {author} {\bibfnamefont {K.}~\bibnamefont {Watanabe}}, \bibinfo {author} {\bibfnamefont {T.}~\bibnamefont {Taniguchi}}, \bibinfo {author} {\bibfnamefont {E.}~\bibnamefont {Kaxiras}},\ and\ \bibinfo {author} {\bibfnamefont {P.}~\bibnamefont {Jarillo-Herrero}},\ }\bibfield  {title} {\bibinfo {title} {Unconventional superconductivity in magic-angle graphene superlattices},\ }\href {https://doi.org/10.1038/nature26160} {\bibfield  {journal} {\bibinfo  {journal} {Nature}\ }\textbf {\bibinfo {volume} {556}},\ \bibinfo {pages} {43} (\bibinfo {year} {2018}{\natexlab{b}})}\BibitemShut {NoStop}%
\bibitem [{\citenamefont {Yankowitz}\ \emph {et~al.}(2019)\citenamefont {Yankowitz}, \citenamefont {Chen}, \citenamefont {Polshyn}, \citenamefont {Zhang}, \citenamefont {Watanabe}, \citenamefont {Taniguchi}, \citenamefont {Graf}, \citenamefont {Young},\ and\ \citenamefont {Dean}}]{Yankowitz2019}%
  \BibitemOpen
  \bibfield  {author} {\bibinfo {author} {\bibfnamefont {M.}~\bibnamefont {Yankowitz}}, \bibinfo {author} {\bibfnamefont {S.}~\bibnamefont {Chen}}, \bibinfo {author} {\bibfnamefont {H.}~\bibnamefont {Polshyn}}, \bibinfo {author} {\bibfnamefont {Y.}~\bibnamefont {Zhang}}, \bibinfo {author} {\bibfnamefont {K.}~\bibnamefont {Watanabe}}, \bibinfo {author} {\bibfnamefont {T.}~\bibnamefont {Taniguchi}}, \bibinfo {author} {\bibfnamefont {D.}~\bibnamefont {Graf}}, \bibinfo {author} {\bibfnamefont {A.~F.}\ \bibnamefont {Young}},\ and\ \bibinfo {author} {\bibfnamefont {C.~R.}\ \bibnamefont {Dean}},\ }\bibfield  {title} {\bibinfo {title} {Tuning superconductivity in twisted bilayer graphene},\ }\href {https://doi.org/10.1126/science.aav1910} {\bibfield  {journal} {\bibinfo  {journal} {Science}\ }\textbf {\bibinfo {volume} {363}},\ \bibinfo {pages} {1059} (\bibinfo {year} {2019})}\BibitemShut {NoStop}%
\bibitem [{\citenamefont {Lu}\ \emph {et~al.}(2019)\citenamefont {Lu}, \citenamefont {Stepanov}, \citenamefont {Yang}, \citenamefont {Xie}, \citenamefont {Aamir}, \citenamefont {Das}, \citenamefont {Urgell}, \citenamefont {Watanabe}, \citenamefont {Taniguchi}, \citenamefont {Zhang}, \citenamefont {Bachtold}, \citenamefont {MacDonald},\ and\ \citenamefont {Efetov}}]{Lu2019}%
  \BibitemOpen
  \bibfield  {author} {\bibinfo {author} {\bibfnamefont {X.}~\bibnamefont {Lu}}, \bibinfo {author} {\bibfnamefont {P.}~\bibnamefont {Stepanov}}, \bibinfo {author} {\bibfnamefont {W.}~\bibnamefont {Yang}}, \bibinfo {author} {\bibfnamefont {M.}~\bibnamefont {Xie}}, \bibinfo {author} {\bibfnamefont {M.~A.}\ \bibnamefont {Aamir}}, \bibinfo {author} {\bibfnamefont {I.}~\bibnamefont {Das}}, \bibinfo {author} {\bibfnamefont {C.}~\bibnamefont {Urgell}}, \bibinfo {author} {\bibfnamefont {K.}~\bibnamefont {Watanabe}}, \bibinfo {author} {\bibfnamefont {T.}~\bibnamefont {Taniguchi}}, \bibinfo {author} {\bibfnamefont {G.}~\bibnamefont {Zhang}}, \bibinfo {author} {\bibfnamefont {A.}~\bibnamefont {Bachtold}}, \bibinfo {author} {\bibfnamefont {A.~H.}\ \bibnamefont {MacDonald}},\ and\ \bibinfo {author} {\bibfnamefont {D.~K.}\ \bibnamefont {Efetov}},\ }\bibfield  {title} {\bibinfo {title} {Superconductors, orbital magnets and correlated states in magic-angle bilayer graphene},\ }\href {https://doi.org/10.1038/s41586-019-1695-0} {\bibfield  {journal} {\bibinfo  {journal} {Nature}\ }\textbf {\bibinfo {volume} {574}},\ \bibinfo {pages} {653} (\bibinfo {year} {2019})}\BibitemShut {NoStop}%
\bibitem [{\citenamefont {Wu}\ \emph {et~al.}(2018)\citenamefont {Wu}, \citenamefont {MacDonald},\ and\ \citenamefont {Martin}}]{WuMacDonaldMartin2018}%
  \BibitemOpen
  \bibfield  {author} {\bibinfo {author} {\bibfnamefont {F.}~\bibnamefont {Wu}}, \bibinfo {author} {\bibfnamefont {A.~H.}\ \bibnamefont {MacDonald}},\ and\ \bibinfo {author} {\bibfnamefont {I.}~\bibnamefont {Martin}},\ }\bibfield  {title} {\bibinfo {title} {Theory of phonon-mediated superconductivity in twisted bilayer graphene},\ }\href {https://doi.org/10.1103/PhysRevLett.121.257001} {\bibfield  {journal} {\bibinfo  {journal} {Phys. Rev. Lett.}\ }\textbf {\bibinfo {volume} {121}},\ \bibinfo {pages} {257001} (\bibinfo {year} {2018})}\BibitemShut {NoStop}%
\bibitem [{\citenamefont {Lian}\ \emph {et~al.}(2019)\citenamefont {Lian}, \citenamefont {Wang},\ and\ \citenamefont {Bernevig}}]{Lian2019}%
  \BibitemOpen
  \bibfield  {author} {\bibinfo {author} {\bibfnamefont {B.}~\bibnamefont {Lian}}, \bibinfo {author} {\bibfnamefont {Z.}~\bibnamefont {Wang}},\ and\ \bibinfo {author} {\bibfnamefont {B.~A.}\ \bibnamefont {Bernevig}},\ }\bibfield  {title} {\bibinfo {title} {Twisted bilayer graphene: A phonon-driven superconductor},\ }\href {https://doi.org/10.1103/PhysRevLett.122.257002} {\bibfield  {journal} {\bibinfo  {journal} {Phys. Rev. Lett.}\ }\textbf {\bibinfo {volume} {122}},\ \bibinfo {pages} {257002} (\bibinfo {year} {2019})}\BibitemShut {NoStop}%
\bibitem [{\citenamefont {Isobe}\ \emph {et~al.}(2018)\citenamefont {Isobe}, \citenamefont {Yuan},\ and\ \citenamefont {Fu}}]{Isobe2018}%
  \BibitemOpen
  \bibfield  {author} {\bibinfo {author} {\bibfnamefont {H.}~\bibnamefont {Isobe}}, \bibinfo {author} {\bibfnamefont {N.~F.~Q.}\ \bibnamefont {Yuan}},\ and\ \bibinfo {author} {\bibfnamefont {L.}~\bibnamefont {Fu}},\ }\bibfield  {title} {\bibinfo {title} {Unconventional superconductivity and density waves in twisted bilayer graphene},\ }\href {https://doi.org/10.1103/PhysRevX.8.041041} {\bibfield  {journal} {\bibinfo  {journal} {Phys. Rev. X}\ }\textbf {\bibinfo {volume} {8}},\ \bibinfo {pages} {041041} (\bibinfo {year} {2018})}\BibitemShut {NoStop}%
\bibitem [{\citenamefont {Kennes}\ \emph {et~al.}(2018)\citenamefont {Kennes}, \citenamefont {Lischner},\ and\ \citenamefont {Karrasch}}]{Kennes2018}%
  \BibitemOpen
  \bibfield  {author} {\bibinfo {author} {\bibfnamefont {D.~M.}\ \bibnamefont {Kennes}}, \bibinfo {author} {\bibfnamefont {J.}~\bibnamefont {Lischner}},\ and\ \bibinfo {author} {\bibfnamefont {C.}~\bibnamefont {Karrasch}},\ }\bibfield  {title} {\bibinfo {title} {Strong correlations and {$d+id$} superconductivity in twisted bilayer graphene},\ }\href {https://doi.org/10.1103/PhysRevB.98.241407} {\bibfield  {journal} {\bibinfo  {journal} {Phys. Rev. B}\ }\textbf {\bibinfo {volume} {98}},\ \bibinfo {pages} {241407(R)} (\bibinfo {year} {2018})}\BibitemShut {NoStop}%
\bibitem [{\citenamefont {Gonz{\'a}lez}\ and\ \citenamefont {Stauber}(2019)}]{Gonzalez2019}%
  \BibitemOpen
  \bibfield  {author} {\bibinfo {author} {\bibfnamefont {J.}~\bibnamefont {Gonz{\'a}lez}}\ and\ \bibinfo {author} {\bibfnamefont {T.}~\bibnamefont {Stauber}},\ }\bibfield  {title} {\bibinfo {title} {{Kohn-Luttinger} superconductivity in twisted bilayer graphene},\ }\href {https://doi.org/10.1103/PhysRevLett.122.026801} {\bibfield  {journal} {\bibinfo  {journal} {Phys. Rev. Lett.}\ }\textbf {\bibinfo {volume} {122}},\ \bibinfo {pages} {026801} (\bibinfo {year} {2019})}\BibitemShut {NoStop}%
\bibitem [{\citenamefont {You}\ and\ \citenamefont {Vishwanath}(2019)}]{YouVishwanath2019}%
  \BibitemOpen
  \bibfield  {author} {\bibinfo {author} {\bibfnamefont {Y.-Z.}\ \bibnamefont {You}}\ and\ \bibinfo {author} {\bibfnamefont {A.}~\bibnamefont {Vishwanath}},\ }\bibfield  {title} {\bibinfo {title} {Superconductivity from valley fluctuations and approximate {SO(4)} symmetry in a weak coupling theory of twisted bilayer graphene},\ }\href {https://doi.org/10.1038/s41535-019-0153-4} {\bibfield  {journal} {\bibinfo  {journal} {npj Quantum Mater.}\ }\textbf {\bibinfo {volume} {4}},\ \bibinfo {pages} {16} (\bibinfo {year} {2019})}\BibitemShut {NoStop}%
\bibitem [{\citenamefont {Oh}\ \emph {et~al.}(2021)\citenamefont {Oh}, \citenamefont {Nuckolls}, \citenamefont {Wong}, \citenamefont {Lee}, \citenamefont {Liu}, \citenamefont {Watanabe}, \citenamefont {Taniguchi},\ and\ \citenamefont {Yazdani}}]{Oh2021STM}%
  \BibitemOpen
  \bibfield  {author} {\bibinfo {author} {\bibfnamefont {M.}~\bibnamefont {Oh}}, \bibinfo {author} {\bibfnamefont {K.~P.}\ \bibnamefont {Nuckolls}}, \bibinfo {author} {\bibfnamefont {D.}~\bibnamefont {Wong}}, \bibinfo {author} {\bibfnamefont {R.~L.}\ \bibnamefont {Lee}}, \bibinfo {author} {\bibfnamefont {X.}~\bibnamefont {Liu}}, \bibinfo {author} {\bibfnamefont {K.}~\bibnamefont {Watanabe}}, \bibinfo {author} {\bibfnamefont {T.}~\bibnamefont {Taniguchi}},\ and\ \bibinfo {author} {\bibfnamefont {A.}~\bibnamefont {Yazdani}},\ }\bibfield  {title} {\bibinfo {title} {Evidence for unconventional superconductivity in twisted bilayer graphene},\ }\href {https://doi.org/10.1038/s41586-021-04121-x} {\bibfield  {journal} {\bibinfo  {journal} {Nature}\ }\textbf {\bibinfo {volume} {600}},\ \bibinfo {pages} {240} (\bibinfo {year} {2021})}\BibitemShut {NoStop}%
\bibitem [{\citenamefont {Park}\ \emph {et~al.}(2026)\citenamefont {Park}, \citenamefont {Sun}, \citenamefont {Watanabe}, \citenamefont {Taniguchi},\ and\ \citenamefont {Jarillo-Herrero}}]{Park2026Tunnel}%
  \BibitemOpen
  \bibfield  {author} {\bibinfo {author} {\bibfnamefont {J.~M.}\ \bibnamefont {Park}}, \bibinfo {author} {\bibfnamefont {S.}~\bibnamefont {Sun}}, \bibinfo {author} {\bibfnamefont {K.}~\bibnamefont {Watanabe}}, \bibinfo {author} {\bibfnamefont {T.}~\bibnamefont {Taniguchi}},\ and\ \bibinfo {author} {\bibfnamefont {P.}~\bibnamefont {Jarillo-Herrero}},\ }\bibfield  {title} {\bibinfo {title} {Experimental evidence for nodal superconducting gap in moir{\'e} graphene},\ }\href {https://doi.org/10.1126/science.adv8376} {\bibfield  {journal} {\bibinfo  {journal} {Science}\ }\textbf {\bibinfo {volume} {391}},\ \bibinfo {pages} {79} (\bibinfo {year} {2026})}\BibitemShut {NoStop}%
\bibitem [{\citenamefont {Chen}\ \emph {et~al.}(2024)\citenamefont {Chen}, \citenamefont {Nuckolls}, \citenamefont {Ding}, \citenamefont {Miao}, \citenamefont {Wong}, \citenamefont {Oh}, \citenamefont {Lee}, \citenamefont {He}, \citenamefont {Peng}, \citenamefont {Pei}, \citenamefont {Li}, \citenamefont {Hao}, \citenamefont {Yan}, \citenamefont {Xiao}, \citenamefont {Gao}, \citenamefont {Li}, \citenamefont {Zhang}, \citenamefont {Liu}, \citenamefont {He}, \citenamefont {Watanabe}, \citenamefont {Taniguchi}, \citenamefont {Jozwiak}, \citenamefont {Bostwick}, \citenamefont {Rotenberg}, \citenamefont {Li}, \citenamefont {Han}, \citenamefont {Pan}, \citenamefont {Liu}, \citenamefont {Dai}, \citenamefont {Liu}, \citenamefont {Bernevig}, \citenamefont {Wang}, \citenamefont {Yazdani},\ and\ \citenamefont {Chen}}]{Chen2024EPC}%
  \BibitemOpen
  \bibfield  {author} {\bibinfo {author} {\bibfnamefont {C.}~\bibnamefont {Chen}}, \bibinfo {author} {\bibfnamefont {K.~P.}\ \bibnamefont {Nuckolls}}, \bibinfo {author} {\bibfnamefont {S.}~\bibnamefont {Ding}}, \bibinfo {author} {\bibfnamefont {W.}~\bibnamefont {Miao}}, \bibinfo {author} {\bibfnamefont {D.}~\bibnamefont {Wong}}, \bibinfo {author} {\bibfnamefont {M.}~\bibnamefont {Oh}}, \bibinfo {author} {\bibfnamefont {R.~L.}\ \bibnamefont {Lee}}, \bibinfo {author} {\bibfnamefont {S.}~\bibnamefont {He}}, \bibinfo {author} {\bibfnamefont {C.}~\bibnamefont {Peng}}, \bibinfo {author} {\bibfnamefont {D.}~\bibnamefont {Pei}}, \bibinfo {author} {\bibfnamefont {Y.}~\bibnamefont {Li}}, \bibinfo {author} {\bibfnamefont {C.}~\bibnamefont {Hao}}, \bibinfo {author} {\bibfnamefont {H.}~\bibnamefont {Yan}}, \bibinfo {author} {\bibfnamefont {H.}~\bibnamefont {Xiao}}, \bibinfo {author} {\bibfnamefont {H.}~\bibnamefont {Gao}}, \bibinfo {author} {\bibfnamefont {Q.}~\bibnamefont {Li}}, \bibinfo {author} {\bibfnamefont {S.}~\bibnamefont {Zhang}}, \bibinfo {author} {\bibfnamefont {J.}~\bibnamefont {Liu}}, \bibinfo {author} {\bibfnamefont {L.}~\bibnamefont {He}}, \bibinfo {author} {\bibfnamefont {K.}~\bibnamefont {Watanabe}}, \bibinfo {author} {\bibfnamefont {T.}~\bibnamefont {Taniguchi}}, \bibinfo {author} {\bibfnamefont {C.}~\bibnamefont {Jozwiak}}, \bibinfo {author} {\bibfnamefont {A.}~\bibnamefont {Bostwick}}, \bibinfo {author} {\bibfnamefont {E.}~\bibnamefont {Rotenberg}}, \bibinfo {author} {\bibfnamefont {C.}~\bibnamefont {Li}}, \bibinfo {author} {\bibfnamefont {X.}~\bibnamefont {Han}}, \bibinfo {author} {\bibfnamefont {D.}~\bibnamefont {Pan}}, \bibinfo {author} {\bibfnamefont {Z.}~\bibnamefont {Liu}}, \bibinfo {author} {\bibfnamefont {X.}~\bibnamefont {Dai}}, \bibinfo {author} {\bibfnamefont {C.}~\bibnamefont {Liu}}, \bibinfo {author} {\bibfnamefont {B.~A.}\ \bibnamefont {Bernevig}}, \bibinfo {author} {\bibfnamefont {Y.}~\bibnamefont {Wang}}, \bibinfo {author} {\bibfnamefont {A.}~\bibnamefont {Yazdani}},\ and\ \bibinfo {author} {\bibfnamefont {Y.}~\bibnamefont {Chen}},\ }\bibfield  {title} {\bibinfo {title} {Strong electron--phonon coupling in magic-angle twisted bilayer graphene},\ }\href {https://doi.org/10.1038/s41586-024-08227-w} {\bibfield  {journal} {\bibinfo  {journal} {Nature}\ }\textbf {\bibinfo {volume} {636}},\ \bibinfo {pages} {342} (\bibinfo {year} {2024})}\BibitemShut {NoStop}%
\bibitem [{\citenamefont {Blason}\ and\ \citenamefont {Fabrizio}(2022)}]{BlasonFabrizio2022}%
  \BibitemOpen
  \bibfield  {author} {\bibinfo {author} {\bibfnamefont {A.}~\bibnamefont {Blason}}\ and\ \bibinfo {author} {\bibfnamefont {M.}~\bibnamefont {Fabrizio}},\ }\bibfield  {title} {\bibinfo {title} {Local {Kekul\'e} distortion turns twisted bilayer graphene into topological {Mott} insulators and superconductors},\ }\href {https://doi.org/10.1103/PhysRevB.106.235112} {\bibfield  {journal} {\bibinfo  {journal} {Phys. Rev. B}\ }\textbf {\bibinfo {volume} {106}},\ \bibinfo {pages} {235112} (\bibinfo {year} {2022})}\BibitemShut {NoStop}%
\bibitem [{\citenamefont {Wang}\ \emph {et~al.}(2024)\citenamefont {Wang}, \citenamefont {Zhou}, \citenamefont {Peng}, \citenamefont {Lian},\ and\ \citenamefont {Song}}]{WangMolecular2024}%
  \BibitemOpen
  \bibfield  {author} {\bibinfo {author} {\bibfnamefont {Y.-J.}\ \bibnamefont {Wang}}, \bibinfo {author} {\bibfnamefont {G.-D.}\ \bibnamefont {Zhou}}, \bibinfo {author} {\bibfnamefont {S.-Y.}\ \bibnamefont {Peng}}, \bibinfo {author} {\bibfnamefont {B.}~\bibnamefont {Lian}},\ and\ \bibinfo {author} {\bibfnamefont {Z.-D.}\ \bibnamefont {Song}},\ }\bibfield  {title} {\bibinfo {title} {Molecular pairing in twisted bilayer graphene superconductivity},\ }\href {https://doi.org/10.1103/PhysRevLett.133.146001} {\bibfield  {journal} {\bibinfo  {journal} {Phys. Rev. Lett.}\ }\textbf {\bibinfo {volume} {133}},\ \bibinfo {pages} {146001} (\bibinfo {year} {2024})}\BibitemShut {NoStop}%
\bibitem [{\citenamefont {Wang}\ \emph {et~al.}(2025)\citenamefont {Wang}, \citenamefont {Zhou}, \citenamefont {Lian},\ and\ \citenamefont {Song}}]{WangEPC2025}%
  \BibitemOpen
  \bibfield  {author} {\bibinfo {author} {\bibfnamefont {Y.-J.}\ \bibnamefont {Wang}}, \bibinfo {author} {\bibfnamefont {G.-D.}\ \bibnamefont {Zhou}}, \bibinfo {author} {\bibfnamefont {B.}~\bibnamefont {Lian}},\ and\ \bibinfo {author} {\bibfnamefont {Z.-D.}\ \bibnamefont {Song}},\ }\bibfield  {title} {\bibinfo {title} {Electron-phonon coupling in the topological heavy fermion model of twisted bilayer graphene},\ }\href {https://doi.org/10.1103/PhysRevB.111.035110} {\bibfield  {journal} {\bibinfo  {journal} {Phys. Rev. B}\ }\textbf {\bibinfo {volume} {111}},\ \bibinfo {pages} {035110} (\bibinfo {year} {2025})}\BibitemShut {NoStop}%
\bibitem [{\citenamefont {Zhao}\ and\ \citenamefont {Zhang}(2026)}]{ZhaoZhang2025}%
  \BibitemOpen
  \bibfield  {author} {\bibinfo {author} {\bibfnamefont {J.-Y.}\ \bibnamefont {Zhao}}\ and\ \bibinfo {author} {\bibfnamefont {Y.-H.}\ \bibnamefont {Zhang}},\ }\bibfield  {title} {\bibinfo {title} {Resonating-valence-bond superconductor from small {Fermi} surface in twisted bilayer graphene},\ }\href {https://doi.org/10.1038/s41467-026-76586-1} {\bibfield  {journal} {\bibinfo  {journal} {Nat. Commun.}\ }\textbf {\bibinfo {volume} {17}},\ \bibinfo {pages} {10076} (\bibinfo {year} {2026})}\BibitemShut {NoStop}%
\bibitem [{\citenamefont {Zhang}\ and\ \citenamefont {Mao}(2020)}]{ZhangMao2020}%
  \BibitemOpen
  \bibfield  {author} {\bibinfo {author} {\bibfnamefont {Y.-H.}\ \bibnamefont {Zhang}}\ and\ \bibinfo {author} {\bibfnamefont {D.}~\bibnamefont {Mao}},\ }\bibfield  {title} {\bibinfo {title} {Spin liquids and pseudogap metals in the {SU(4)} {Hubbard} model in a moir\'e superlattice},\ }\href {https://doi.org/10.1103/PhysRevB.101.035122} {\bibfield  {journal} {\bibinfo  {journal} {Phys. Rev. B}\ }\textbf {\bibinfo {volume} {101}},\ \bibinfo {pages} {035122} (\bibinfo {year} {2020})}\BibitemShut {NoStop}%
\bibitem [{\citenamefont {Yang}\ \emph {et~al.}(2024)\citenamefont {Yang}, \citenamefont {Oh},\ and\ \citenamefont {Zhang}}]{YangOhZhang2024}%
  \BibitemOpen
  \bibfield  {author} {\bibinfo {author} {\bibfnamefont {H.}~\bibnamefont {Yang}}, \bibinfo {author} {\bibfnamefont {H.}~\bibnamefont {Oh}},\ and\ \bibinfo {author} {\bibfnamefont {Y.-H.}\ \bibnamefont {Zhang}},\ }\bibfield  {title} {\bibinfo {title} {Strong pairing from a small {F}ermi surface beyond weak coupling: Application to {La$_3$Ni$_2$O$_7$}},\ }\href {https://doi.org/10.1103/PhysRevB.110.104517} {\bibfield  {journal} {\bibinfo  {journal} {Phys. Rev. B}\ }\textbf {\bibinfo {volume} {110}},\ \bibinfo {pages} {104517} (\bibinfo {year} {2024})}\BibitemShut {NoStop}%
\bibitem [{\citenamefont {Wu}\ \emph {et~al.}(2024)\citenamefont {Wu}, \citenamefont {Yang},\ and\ \citenamefont {Zhang}}]{WuYangZhang2024}%
  \BibitemOpen
  \bibfield  {author} {\bibinfo {author} {\bibfnamefont {X.}~\bibnamefont {Wu}}, \bibinfo {author} {\bibfnamefont {H.}~\bibnamefont {Yang}},\ and\ \bibinfo {author} {\bibfnamefont {Y.-H.}\ \bibnamefont {Zhang}},\ }\bibfield  {title} {\bibinfo {title} {Deconfined {Fermi} liquid to {Fermi} liquid transition and superconducting instability},\ }\href {https://doi.org/10.1103/PhysRevB.110.125122} {\bibfield  {journal} {\bibinfo  {journal} {Phys. Rev. B}\ }\textbf {\bibinfo {volume} {110}},\ \bibinfo {pages} {125122} (\bibinfo {year} {2024})}\BibitemShut {NoStop}%
\bibitem [{\citenamefont {White}(1992)}]{White1992}%
  \BibitemOpen
  \bibfield  {author} {\bibinfo {author} {\bibfnamefont {S.~R.}\ \bibnamefont {White}},\ }\bibfield  {title} {\bibinfo {title} {Density matrix formulation for quantum renormalization groups},\ }\href {https://doi.org/10.1103/PhysRevLett.69.2863} {\bibfield  {journal} {\bibinfo  {journal} {Phys. Rev. Lett.}\ }\textbf {\bibinfo {volume} {69}},\ \bibinfo {pages} {2863} (\bibinfo {year} {1992})}\BibitemShut {NoStop}%
\bibitem [{SM()}]{SM}%
  \BibitemOpen
  \href@noop {} {}\bibinfo {note} {See Supplemental Material at [URL will be inserted by publisher] for the reduction of the two-orbital Hubbard model to the $t$--$J_A$ model, the derivation of the effective model and the six pivot amplitudes $g_\eta$, the exact solution of the two-carrier problem (binding threshold, binding energy, effective mass, and pair size), the Gutzwiller approximation and the mean-field equations, the momentum distribution $n(\mathbf k)$ at large $J_A$, and correlation lengths with the intersite flavor exchange $J_\parallel$.}\BibitemShut {Stop}%
\bibitem [{\citenamefont {White}(1993)}]{White1993}%
  \BibitemOpen
  \bibfield  {author} {\bibinfo {author} {\bibfnamefont {S.~R.}\ \bibnamefont {White}},\ }\bibfield  {title} {\bibinfo {title} {Density-matrix algorithms for quantum renormalization groups},\ }\href {https://doi.org/10.1103/PhysRevB.48.10345} {\bibfield  {journal} {\bibinfo  {journal} {Phys. Rev. B}\ }\textbf {\bibinfo {volume} {48}},\ \bibinfo {pages} {10345} (\bibinfo {year} {1993})}\BibitemShut {NoStop}%
\bibitem [{\citenamefont {Hauschild}\ and\ \citenamefont {Pollmann}(2018)}]{Hauschild2018}%
  \BibitemOpen
  \bibfield  {author} {\bibinfo {author} {\bibfnamefont {J.}~\bibnamefont {Hauschild}}\ and\ \bibinfo {author} {\bibfnamefont {F.}~\bibnamefont {Pollmann}},\ }\bibfield  {title} {\bibinfo {title} {Efficient numerical simulations with tensor networks: Tensor network {P}ython ({TeNPy})},\ }\bibfield  {journal} {\bibinfo  {journal} {SciPost Phys. Lect. Notes}\ }\textbf {\bibinfo {volume} {5}},\ \href {https://doi.org/10.21468/SciPostPhysLectNotes.5} {10.21468/SciPostPhysLectNotes.5} (\bibinfo {year} {2018})\BibitemShut {NoStop}%
\bibitem [{\citenamefont {Hauschild}\ \emph {et~al.}(2024)\citenamefont {Hauschild}, \citenamefont {Unfried}, \citenamefont {Anand}, \citenamefont {Andrews}, \citenamefont {Bintz}, \citenamefont {Borla}, \citenamefont {Divic}, \citenamefont {Drescher}, \citenamefont {Geiger}, \citenamefont {Hefel}, \citenamefont {H{\'e}mery}, \citenamefont {Kadow}, \citenamefont {Kemp}, \citenamefont {Kirchner}, \citenamefont {Liu}, \citenamefont {M{\"o}ller}, \citenamefont {Parker}, \citenamefont {Rader}, \citenamefont {Romen}, \citenamefont {Scalet}, \citenamefont {Schoonderwoerd}, \citenamefont {Schulz}, \citenamefont {Soejima}, \citenamefont {Thoma}, \citenamefont {Wu}, \citenamefont {Zechmann}, \citenamefont {Zweng}, \citenamefont {Mong}, \citenamefont {Zaletel},\ and\ \citenamefont {Pollmann}}]{Hauschild2024}%
  \BibitemOpen
  \bibfield  {author} {\bibinfo {author} {\bibfnamefont {J.}~\bibnamefont {Hauschild}}, \bibinfo {author} {\bibfnamefont {J.}~\bibnamefont {Unfried}}, \bibinfo {author} {\bibfnamefont {S.}~\bibnamefont {Anand}}, \bibinfo {author} {\bibfnamefont {B.}~\bibnamefont {Andrews}}, \bibinfo {author} {\bibfnamefont {M.}~\bibnamefont {Bintz}}, \bibinfo {author} {\bibfnamefont {U.}~\bibnamefont {Borla}}, \bibinfo {author} {\bibfnamefont {S.}~\bibnamefont {Divic}}, \bibinfo {author} {\bibfnamefont {M.}~\bibnamefont {Drescher}}, \bibinfo {author} {\bibfnamefont {J.}~\bibnamefont {Geiger}}, \bibinfo {author} {\bibfnamefont {M.}~\bibnamefont {Hefel}}, \bibinfo {author} {\bibfnamefont {K.}~\bibnamefont {H{\'e}mery}}, \bibinfo {author} {\bibfnamefont {W.}~\bibnamefont {Kadow}}, \bibinfo {author} {\bibfnamefont {J.}~\bibnamefont {Kemp}}, \bibinfo {author} {\bibfnamefont {N.}~\bibnamefont {Kirchner}}, \bibinfo {author} {\bibfnamefont {V.~S.}\ \bibnamefont {Liu}}, \bibinfo {author} {\bibfnamefont {G.}~\bibnamefont {M{\"o}ller}}, \bibinfo {author} {\bibfnamefont {D.}~\bibnamefont {Parker}}, \bibinfo {author} {\bibfnamefont {M.}~\bibnamefont {Rader}}, \bibinfo {author} {\bibfnamefont {A.}~\bibnamefont {Romen}}, \bibinfo {author} {\bibfnamefont {S.}~\bibnamefont {Scalet}}, \bibinfo {author} {\bibfnamefont {L.}~\bibnamefont {Schoonderwoerd}}, \bibinfo {author} {\bibfnamefont {M.}~\bibnamefont {Schulz}}, \bibinfo {author} {\bibfnamefont {T.}~\bibnamefont {Soejima}}, \bibinfo {author} {\bibfnamefont {P.}~\bibnamefont {Thoma}}, \bibinfo {author} {\bibfnamefont {Y.}~\bibnamefont {Wu}}, \bibinfo {author} {\bibfnamefont {P.}~\bibnamefont {Zechmann}}, \bibinfo {author} {\bibfnamefont {L.}~\bibnamefont {Zweng}}, \bibinfo {author} {\bibfnamefont {R.}~\bibnamefont {Mong}}, \bibinfo {author} {\bibfnamefont {M.~P.}\ \bibnamefont {Zaletel}},\ and\ \bibinfo {author} {\bibfnamefont {F.}~\bibnamefont {Pollmann}},\ }\bibfield  {title} {\bibinfo {title} {Tensor network {P}ython ({TeNPy}) version 1},\ }\bibfield  {journal} {\bibinfo  {journal} {SciPost Phys. Codebases}\ }\textbf {\bibinfo {volume} {41}},\ \href {https://doi.org/10.21468/SciPostPhysCodeb.41} {10.21468/SciPostPhysCodeb.41} (\bibinfo {year} {2024})\BibitemShut {NoStop}%
\bibitem [{\citenamefont {Miyakawa}\ \emph {et~al.}(1998)\citenamefont {Miyakawa}, \citenamefont {Guptasarma}, \citenamefont {Zasadzinski}, \citenamefont {Hinks},\ and\ \citenamefont {Gray}}]{Miyakawa1998}%
  \BibitemOpen
  \bibfield  {author} {\bibinfo {author} {\bibfnamefont {N.}~\bibnamefont {Miyakawa}}, \bibinfo {author} {\bibfnamefont {P.}~\bibnamefont {Guptasarma}}, \bibinfo {author} {\bibfnamefont {J.~F.}\ \bibnamefont {Zasadzinski}}, \bibinfo {author} {\bibfnamefont {D.~G.}\ \bibnamefont {Hinks}},\ and\ \bibinfo {author} {\bibfnamefont {K.~E.}\ \bibnamefont {Gray}},\ }\bibfield  {title} {\bibinfo {title} {Strong dependence of the superconducting gap on oxygen doping from tunneling measurements on {Bi$_2$Sr$_2$CaCu$_2$O$_{8-\delta}$}},\ }\href {https://doi.org/10.1103/PhysRevLett.80.157} {\bibfield  {journal} {\bibinfo  {journal} {Phys. Rev. Lett.}\ }\textbf {\bibinfo {volume} {80}},\ \bibinfo {pages} {157} (\bibinfo {year} {1998})}\BibitemShut {NoStop}%
\bibitem [{\citenamefont {Lee}\ \emph {et~al.}(2006)\citenamefont {Lee}, \citenamefont {Nagaosa},\ and\ \citenamefont {Wen}}]{LeeNagaosaWen2006}%
  \BibitemOpen
  \bibfield  {author} {\bibinfo {author} {\bibfnamefont {P.~A.}\ \bibnamefont {Lee}}, \bibinfo {author} {\bibfnamefont {N.}~\bibnamefont {Nagaosa}},\ and\ \bibinfo {author} {\bibfnamefont {X.-G.}\ \bibnamefont {Wen}},\ }\bibfield  {title} {\bibinfo {title} {Doping a {Mott} insulator: Physics of high-temperature superconductivity},\ }\href {https://doi.org/10.1103/RevModPhys.78.17} {\bibfield  {journal} {\bibinfo  {journal} {Rev. Mod. Phys.}\ }\textbf {\bibinfo {volume} {78}},\ \bibinfo {pages} {17} (\bibinfo {year} {2006})}\BibitemShut {NoStop}%
\bibitem [{\citenamefont {Oh}\ \emph {et~al.}(2026)\citenamefont {Oh}, \citenamefont {Yang},\ and\ \citenamefont {Zhang}}]{OhYangZhangKineticSC}%
  \BibitemOpen
  \bibfield  {author} {\bibinfo {author} {\bibfnamefont {H.}~\bibnamefont {Oh}}, \bibinfo {author} {\bibfnamefont {H.}~\bibnamefont {Yang}},\ and\ \bibinfo {author} {\bibfnamefont {Y.-H.}\ \bibnamefont {Zhang}},\ }\bibfield  {title} {\bibinfo {title} {High-temperature superconductivity from kinetic energy},\ }\href {https://doi.org/10.1103/xrjk-y28y} {\bibfield  {journal} {\bibinfo  {journal} {Phys. Rev. B}\ }\textbf {\bibinfo {volume} {114}},\ \bibinfo {pages} {274103} (\bibinfo {year} {2026})}\BibitemShut {NoStop}%
\bibitem [{\citenamefont {McCulloch}(2008)}]{McCulloch2008}%
  \BibitemOpen
  \bibfield  {author} {\bibinfo {author} {\bibfnamefont {I.~P.}\ \bibnamefont {McCulloch}},\ }\href@noop {} {\bibinfo {title} {Infinite size density matrix renormalization group, revisited}} (\bibinfo {year} {2008}),\ \Eprint {https://arxiv.org/abs/0804.2509} {arXiv:0804.2509 [cond-mat.str-el]} \BibitemShut {NoStop}%
\bibitem [{\citenamefont {Jiang}\ \emph {et~al.}(2018)\citenamefont {Jiang}, \citenamefont {Weng},\ and\ \citenamefont {Kivelson}}]{JiangWengKivelson2018}%
  \BibitemOpen
  \bibfield  {author} {\bibinfo {author} {\bibfnamefont {H.-C.}\ \bibnamefont {Jiang}}, \bibinfo {author} {\bibfnamefont {Z.-Y.}\ \bibnamefont {Weng}},\ and\ \bibinfo {author} {\bibfnamefont {S.~A.}\ \bibnamefont {Kivelson}},\ }\bibfield  {title} {\bibinfo {title} {Superconductivity in the doped {$t$}-{$J$} model: Results for four-leg cylinders},\ }\href {https://doi.org/10.1103/PhysRevB.98.140505} {\bibfield  {journal} {\bibinfo  {journal} {Phys. Rev. B}\ }\textbf {\bibinfo {volume} {98}},\ \bibinfo {pages} {140505(R)} (\bibinfo {year} {2018})}\BibitemShut {NoStop}%
\bibitem [{\citenamefont {Chao}\ \emph {et~al.}(1977)\citenamefont {Chao}, \citenamefont {Spa{\l}ek},\ and\ \citenamefont {Ole{\'s}}}]{ChaoSpalekOles1977}%
  \BibitemOpen
  \bibfield  {author} {\bibinfo {author} {\bibfnamefont {K.~A.}\ \bibnamefont {Chao}}, \bibinfo {author} {\bibfnamefont {J.}~\bibnamefont {Spa{\l}ek}},\ and\ \bibinfo {author} {\bibfnamefont {A.~M.}\ \bibnamefont {Ole{\'s}}},\ }\bibfield  {title} {\bibinfo {title} {Kinetic exchange interaction in a narrow {S}-band},\ }\href {https://doi.org/10.1088/0022-3719/10/10/002} {\bibfield  {journal} {\bibinfo  {journal} {J. Phys. C: Solid State Phys.}\ }\textbf {\bibinfo {volume} {10}},\ \bibinfo {pages} {L271} (\bibinfo {year} {1977})}\BibitemShut {NoStop}%
\bibitem [{\citenamefont {MacDonald}\ \emph {et~al.}(1988)\citenamefont {MacDonald}, \citenamefont {Girvin},\ and\ \citenamefont {Yoshioka}}]{MacDonaldGirvinYoshioka1988}%
  \BibitemOpen
  \bibfield  {author} {\bibinfo {author} {\bibfnamefont {A.~H.}\ \bibnamefont {MacDonald}}, \bibinfo {author} {\bibfnamefont {S.~M.}\ \bibnamefont {Girvin}},\ and\ \bibinfo {author} {\bibfnamefont {D.}~\bibnamefont {Yoshioka}},\ }\bibfield  {title} {\bibinfo {title} {{$t/U$} expansion for the {Hubbard} model},\ }\href {https://doi.org/10.1103/PhysRevB.37.9753} {\bibfield  {journal} {\bibinfo  {journal} {Phys. Rev. B}\ }\textbf {\bibinfo {volume} {37}},\ \bibinfo {pages} {9753} (\bibinfo {year} {1988})}\BibitemShut {NoStop}%
\bibitem [{\citenamefont {Cr{\'e}pel}\ and\ \citenamefont {Fu}(2021)}]{CrepelFu2021}%
  \BibitemOpen
  \bibfield  {author} {\bibinfo {author} {\bibfnamefont {V.}~\bibnamefont {Cr{\'e}pel}}\ and\ \bibinfo {author} {\bibfnamefont {L.}~\bibnamefont {Fu}},\ }\bibfield  {title} {\bibinfo {title} {New mechanism and exact theory of superconductivity from strong repulsive interaction},\ }\href {https://doi.org/10.1126/sciadv.abh2233} {\bibfield  {journal} {\bibinfo  {journal} {Sci. Adv.}\ }\textbf {\bibinfo {volume} {7}},\ \bibinfo {pages} {eabh2233} (\bibinfo {year} {2021})}\BibitemShut {NoStop}%
\bibitem [{\citenamefont {Chien}\ \emph {et~al.}(2008)\citenamefont {Chien}, \citenamefont {Chen},\ and\ \citenamefont {Levin}}]{ChienChenLevin2008}%
  \BibitemOpen
  \bibfield  {author} {\bibinfo {author} {\bibfnamefont {C.-C.}\ \bibnamefont {Chien}}, \bibinfo {author} {\bibfnamefont {Q.}~\bibnamefont {Chen}},\ and\ \bibinfo {author} {\bibfnamefont {K.}~\bibnamefont {Levin}},\ }\bibfield  {title} {\bibinfo {title} {Fermions with attractive interactions on optical lattices and implications for correlated systems},\ }\href {https://doi.org/10.1103/PhysRevA.78.043612} {\bibfield  {journal} {\bibinfo  {journal} {Phys. Rev. A}\ }\textbf {\bibinfo {volume} {78}},\ \bibinfo {pages} {043612} (\bibinfo {year} {2008})}\BibitemShut {NoStop}%
\bibitem [{\citenamefont {Zhang}\ \emph {et~al.}(1988)\citenamefont {Zhang}, \citenamefont {Gros}, \citenamefont {Rice},\ and\ \citenamefont {Shiba}}]{ZhangGrosRiceShiba1988}%
  \BibitemOpen
  \bibfield  {author} {\bibinfo {author} {\bibfnamefont {F.~C.}\ \bibnamefont {Zhang}}, \bibinfo {author} {\bibfnamefont {C.}~\bibnamefont {Gros}}, \bibinfo {author} {\bibfnamefont {T.~M.}\ \bibnamefont {Rice}},\ and\ \bibinfo {author} {\bibfnamefont {H.}~\bibnamefont {Shiba}},\ }\bibfield  {title} {\bibinfo {title} {A renormalised {H}amiltonian approach to a resonant valence bond wavefunction},\ }\href {https://doi.org/10.1088/0953-2048/1/1/009} {\bibfield  {journal} {\bibinfo  {journal} {Supercond. Sci. Technol.}\ }\textbf {\bibinfo {volume} {1}},\ \bibinfo {pages} {36} (\bibinfo {year} {1988})}\BibitemShut {NoStop}%
\bibitem [{\citenamefont {Ledwith}\ \emph {et~al.}(2025)\citenamefont {Ledwith}, \citenamefont {Dong}, \citenamefont {Vishwanath},\ and\ \citenamefont {Khalaf}}]{Ledwith2025}%
  \BibitemOpen
  \bibfield  {author} {\bibinfo {author} {\bibfnamefont {P.~J.}\ \bibnamefont {Ledwith}}, \bibinfo {author} {\bibfnamefont {J.}~\bibnamefont {Dong}}, \bibinfo {author} {\bibfnamefont {A.}~\bibnamefont {Vishwanath}},\ and\ \bibinfo {author} {\bibfnamefont {E.}~\bibnamefont {Khalaf}},\ }\bibfield  {title} {\bibinfo {title} {Nonlocal moments and {Mott} semimetal in the {Chern} bands of twisted bilayer graphene},\ }\href {https://doi.org/10.1103/PhysRevX.15.021087} {\bibfield  {journal} {\bibinfo  {journal} {Phys. Rev. X}\ }\textbf {\bibinfo {volume} {15}},\ \bibinfo {pages} {021087} (\bibinfo {year} {2025})}\BibitemShut {NoStop}%
\bibitem [{\citenamefont {Zhao}\ \emph {et~al.}(2025)\citenamefont {Zhao}, \citenamefont {Zhou},\ and\ \citenamefont {Zhang}}]{ZhaoZhouZhang2025}%
  \BibitemOpen
  \bibfield  {author} {\bibinfo {author} {\bibfnamefont {J.-Y.}\ \bibnamefont {Zhao}}, \bibinfo {author} {\bibfnamefont {B.}~\bibnamefont {Zhou}},\ and\ \bibinfo {author} {\bibfnamefont {Y.-H.}\ \bibnamefont {Zhang}},\ }\bibfield  {title} {\bibinfo {title} {Topological {Mott} localization and pseudogap metal in twisted bilayer graphene},\ }\href {https://doi.org/10.1103/9n8v-7rx2} {\bibfield  {journal} {\bibinfo  {journal} {Phys. Rev. B}\ }\textbf {\bibinfo {volume} {112}},\ \bibinfo {pages} {085107} (\bibinfo {year} {2025})}\BibitemShut {NoStop}%
\bibitem [{\citenamefont {Kwan}\ \emph {et~al.}(2021)\citenamefont {Kwan}, \citenamefont {Wagner}, \citenamefont {Soejima}, \citenamefont {Zaletel}, \citenamefont {Simon}, \citenamefont {Parameswaran},\ and\ \citenamefont {Bultinck}}]{Kwan2021}%
  \BibitemOpen
  \bibfield  {author} {\bibinfo {author} {\bibfnamefont {Y.~H.}\ \bibnamefont {Kwan}}, \bibinfo {author} {\bibfnamefont {G.}~\bibnamefont {Wagner}}, \bibinfo {author} {\bibfnamefont {T.}~\bibnamefont {Soejima}}, \bibinfo {author} {\bibfnamefont {M.~P.}\ \bibnamefont {Zaletel}}, \bibinfo {author} {\bibfnamefont {S.~H.}\ \bibnamefont {Simon}}, \bibinfo {author} {\bibfnamefont {S.~A.}\ \bibnamefont {Parameswaran}},\ and\ \bibinfo {author} {\bibfnamefont {N.}~\bibnamefont {Bultinck}},\ }\bibfield  {title} {\bibinfo {title} {Kekul{\'e} spiral order at all nonzero integer fillings in twisted bilayer graphene},\ }\href {https://doi.org/10.1103/PhysRevX.11.041063} {\bibfield  {journal} {\bibinfo  {journal} {Phys. Rev. X}\ }\textbf {\bibinfo {volume} {11}},\ \bibinfo {pages} {041063} (\bibinfo {year} {2021})}\BibitemShut {NoStop}%
\bibitem [{\citenamefont {Nuckolls}\ \emph {et~al.}(2023)\citenamefont {Nuckolls}, \citenamefont {Lee}, \citenamefont {Oh}, \citenamefont {Wong}, \citenamefont {Soejima}, \citenamefont {Hong}, \citenamefont {C{\u{a}}lug{\u{a}}ru}, \citenamefont {Herzog-Arbeitman}, \citenamefont {Bernevig}, \citenamefont {Watanabe}, \citenamefont {Taniguchi}, \citenamefont {Regnault}, \citenamefont {Zaletel},\ and\ \citenamefont {Yazdani}}]{Nuckolls2023}%
  \BibitemOpen
  \bibfield  {author} {\bibinfo {author} {\bibfnamefont {K.~P.}\ \bibnamefont {Nuckolls}}, \bibinfo {author} {\bibfnamefont {R.~L.}\ \bibnamefont {Lee}}, \bibinfo {author} {\bibfnamefont {M.}~\bibnamefont {Oh}}, \bibinfo {author} {\bibfnamefont {D.}~\bibnamefont {Wong}}, \bibinfo {author} {\bibfnamefont {T.}~\bibnamefont {Soejima}}, \bibinfo {author} {\bibfnamefont {J.~P.}\ \bibnamefont {Hong}}, \bibinfo {author} {\bibfnamefont {D.}~\bibnamefont {C{\u{a}}lug{\u{a}}ru}}, \bibinfo {author} {\bibfnamefont {J.}~\bibnamefont {Herzog-Arbeitman}}, \bibinfo {author} {\bibfnamefont {B.~A.}\ \bibnamefont {Bernevig}}, \bibinfo {author} {\bibfnamefont {K.}~\bibnamefont {Watanabe}}, \bibinfo {author} {\bibfnamefont {T.}~\bibnamefont {Taniguchi}}, \bibinfo {author} {\bibfnamefont {N.}~\bibnamefont {Regnault}}, \bibinfo {author} {\bibfnamefont {M.~P.}\ \bibnamefont {Zaletel}},\ and\ \bibinfo {author} {\bibfnamefont {A.}~\bibnamefont {Yazdani}},\ }\bibfield  {title} {\bibinfo {title} {Quantum textures of the many-body wavefunctions in magic-angle graphene},\ }\href {https://doi.org/10.1038/s41586-023-06226-x} {\bibfield  {journal} {\bibinfo  {journal} {Nature}\ }\textbf {\bibinfo {volume} {620}},\ \bibinfo {pages} {525} (\bibinfo {year} {2023})}\BibitemShut {NoStop}%
\bibitem [{\citenamefont {Kim}\ \emph {et~al.}(2023)\citenamefont {Kim}, \citenamefont {Choi}, \citenamefont {Lantagne-Hurtubise}, \citenamefont {Lewandowski}, \citenamefont {Thomson}, \citenamefont {Kong}, \citenamefont {Zhou}, \citenamefont {Baum}, \citenamefont {Zhang}, \citenamefont {Holleis}, \citenamefont {Watanabe}, \citenamefont {Taniguchi}, \citenamefont {Young}, \citenamefont {Alicea},\ and\ \citenamefont {Nadj-Perge}}]{Kim2023IVC}%
  \BibitemOpen
  \bibfield  {author} {\bibinfo {author} {\bibfnamefont {H.}~\bibnamefont {Kim}}, \bibinfo {author} {\bibfnamefont {Y.}~\bibnamefont {Choi}}, \bibinfo {author} {\bibfnamefont {{\'E}.}~\bibnamefont {Lantagne-Hurtubise}}, \bibinfo {author} {\bibfnamefont {C.}~\bibnamefont {Lewandowski}}, \bibinfo {author} {\bibfnamefont {A.}~\bibnamefont {Thomson}}, \bibinfo {author} {\bibfnamefont {L.}~\bibnamefont {Kong}}, \bibinfo {author} {\bibfnamefont {H.}~\bibnamefont {Zhou}}, \bibinfo {author} {\bibfnamefont {E.}~\bibnamefont {Baum}}, \bibinfo {author} {\bibfnamefont {Y.}~\bibnamefont {Zhang}}, \bibinfo {author} {\bibfnamefont {L.}~\bibnamefont {Holleis}}, \bibinfo {author} {\bibfnamefont {K.}~\bibnamefont {Watanabe}}, \bibinfo {author} {\bibfnamefont {T.}~\bibnamefont {Taniguchi}}, \bibinfo {author} {\bibfnamefont {A.~F.}\ \bibnamefont {Young}}, \bibinfo {author} {\bibfnamefont {J.}~\bibnamefont {Alicea}},\ and\ \bibinfo {author} {\bibfnamefont {S.}~\bibnamefont {Nadj-Perge}},\ }\bibfield  {title} {\bibinfo {title} {Imaging inter-valley coherent order in magic-angle twisted trilayer graphene},\ }\href {https://doi.org/10.1038/s41586-023-06663-8} {\bibfield  {journal} {\bibinfo  {journal} {Nature}\ }\textbf {\bibinfo {volume} {623}},\ \bibinfo {pages} {942} (\bibinfo {year} {2023})}\BibitemShut {NoStop}%
\end{thebibliography}%

\clearpage
\onecolumngrid
\begin{center}
{\large\bfseries Supplemental Material for\\[2pt]
``Pairing from anti-Hund's coupling under infinite Hubbard repulsion''}
\end{center}
\setcounter{equation}{0}
\renewcommand{\theequation}{S\arabic{equation}}
\renewcommand{\theHequation}{S\arabic{equation}}
\setcounter{table}{0}
\renewcommand{\thetable}{S\arabic{table}}
\renewcommand{\theHtable}{S\arabic{table}}
\setcounter{figure}{0}
\renewcommand{\thefigure}{S\arabic{figure}}
\renewcommand{\theHfigure}{S\arabic{figure}}
\setcounter{section}{0}
\renewcommand{\thesection}{\Alph{section}}
\renewcommand{\thesubsection}{\thesection.\arabic{subsection}}
\makeatletter\renewcommand{\p@subsection}{}\makeatother
\renewcommand{\theHsection}{SM.\Alph{section}}
\renewcommand{\theHsubsection}{SM.\Alph{section}.\arabic{subsection}}
\setcounter{secnumdepth}{2}

This Supplemental Material is organized as follows.  We first show how the two-orbital Hubbard model reduces to the four-flavor $t$--$J_A$ model in the limit $U\to\infty$.  We then derive the effective Hamiltonian $\hat H_{\rm eff}$ for $J_A\gg|t|$ to second order in $t/J_A$ and solve it exactly for two carriers.  At finite carrier density, we handle the hard-core constraint in the Gutzwiller approximation and treat the resulting Hamiltonian in mean-field theory, with the on-site pair amplitude constrained to zero.  We present the resulting pairing threshold, single-particle gap, BEC--BCS crossover, and tunneling spectra.  Finally, we provide additional DMRG results on the momentum distribution at large $J_A$ and on the effect of the intersite flavor exchange $J_\parallel$.

\section{From the two-orbital Hubbard model to the $t$--$J_A$ model}
\label{sm:hubbard}

We start with the two-orbital Hubbard model on the triangular lattice,
\begin{equation}
H_{\rm Hub}=-t\sum_{\langle ij\rangle,\alpha}\bigl(c^\dagger_{i\alpha}c_{j\alpha}+{\rm H.c.}\bigr)+\frac U2\sum_in_i^2+J_A\sum_i\mathbf S_{iK}\cdot\mathbf S_{iK'}.
\label{eq:hub}
\end{equation}
Here $\alpha=(\tau,\sigma)$ labels the four flavors, with the valley $\tau=K,K'$ and the spin $\sigma=\uparrow,\downarrow$; $n_i=\sum_\alpha c^\dagger_{i\alpha}c_{i\alpha}$ is the density at site $i$; and $\mathbf S_{i\tau}$ is the spin operator of valley $\tau$. We focus on the regime $U\gg|t|,J_A$ in this work. The electron density is $n=2-x$ per site. Note that $n=2$ corresponds to the insulator, and we consider the regime $1\le n\le2$. The restricted Hilbert space due to the large $U$ consists of four singlon states (with $n_i=1$) and six doublon states (with $n_i=2$) at each site, as shown in Fig.~\ref{fig:model}(a). The empty state is forbidden, since creating an empty site from two singly occupied sites costs energy $U$. Similarly, creating a triply occupied site from two doubly occupied sites costs energy $U$. The Hamiltonian in this restricted Hilbert space becomes Eq.~(\ref{eq:H})~\cite{ZhangMao2020}, where $P_i$ in $\tilde c_{i\alpha}=P_ic_{i\alpha}P_i$ is the projection operator onto the singlon and doublon states at site $i$. The Hubbard term is a constant at fixed density, and the $J_A$ term is unchanged. Virtual hopping to the forbidden states generates additional terms of order $t^2/U$, such as the superexchange, which vanish in the limit $U\to\infty$ considered here.

\section{Effective Hamiltonian to second order in $t/J_A$}
\label{sm:heff}

In this section we derive the effective Hamiltonian $\hat H_{\rm eff}$ of Eq.~(\ref{eq:heffmain}). We split the Hamiltonian of Eq.~(\ref{eq:H}) as $H=H_J+H_t$, with $H_J=J_A\sum_i\mathbf S_{iK}\cdot\mathbf S_{iK'}$ and $H_t=-t\sum_{\langle ij\rangle\alpha}(\tilde c^\dagger_{i\alpha}\tilde c_{j\alpha}+{\rm H.c.})$, and treat $H_t$ as a perturbation for $J_A\gg|t|$. On a doubly occupied site the eigenstate of $H_J$ with the lowest energy, $-\tfrac34J_A$, is the intervalley spin singlet $|s\rangle=(|{\uparrow_K}{\downarrow_{K'}}\rangle-|{\downarrow_K}{\uparrow_{K'}}\rangle)/\sqrt2$. The other five doublons lie higher in energy. The three intervalley spin triplets, $|t_+\rangle=|{\uparrow_K}{\uparrow_{K'}}\rangle$, $|t_0\rangle=(|{\uparrow_K}{\downarrow_{K'}}\rangle+|{\downarrow_K}{\uparrow_{K'}}\rangle)/\sqrt2$, and $|t_-\rangle=|{\downarrow_K}{\downarrow_{K'}}\rangle$, lie at $+\tfrac14J_A$. The two intravalley doublons $|{\uparrow_K}{\downarrow_K}\rangle$ and $|{\uparrow_{K'}}{\downarrow_{K'}}\rangle$ lie at $0$. A singly occupied site, a singlon, has $H_J=0$ for each of its four flavors. At $x=0$ the ground state of $H_J$ is thus the product of on-site singlets, which we take as the vacuum. Each doped hole turns the singlet on one site into a singlon, which we call a carrier. At low energy each site is either $|s\rangle$ or a carrier $|\beta\rangle$ of flavor $\beta$, so the local space has $1+4$ states.

The operator $d^\dagger_{i\beta}=c^\dagger_{i\beta}\hat s_i$ creates a carrier: it replaces $|s\rangle_i$ by $|\beta\rangle_i$, where $\hat s_i$ annihilates the singlet on site $i$. The carrier is a hard-core fermion: two carriers cannot share a site. The carrier number $n^d_i=\sum_\beta d^\dagger_{i\beta}d_{i\beta}$ is $1$ on a carrier and $0$ on a singlet. Within the low-energy space, $d^\dagger_{i\beta}=\pm\sqrt2\,\tilde c_{i\bar\beta}$, where $\bar\beta$ is the partner of $\beta$ in $|s\rangle$. Hence a carrier is a doped hole, and a pair of carriers has charge $2e$.

To second order in $t/J_A$, the carriers are described by the effective Hamiltonian
\begin{equation}
\hat H_{\rm eff}=\frac t2\sum_{\langle ij\rangle\beta}\bigl(d^\dagger_{i\beta}d_{j\beta}+{\rm H.c.}\bigr)+V_{\rm NN}\sum_{\langle ij\rangle}n^d_in^d_j+\hat H_{\rm piv}.
\label{eq:heff}
\end{equation}
Its three terms are the carrier hopping, a nearest-neighbor repulsion with $V_{\rm NN}=+\tfrac{17}{6}\,t^2/J_A$, and the pivot term
\begin{equation}
\hat H_{\rm piv}=\sum_\eta g_\eta\sum_{w\in\eta}\sum_p\sum_{\substack{q,j\,\sim\,p\\ j\neq q}}B^\dagger_{w,pj}B^{\phantom\dagger}_{w,pq},\qquad B^\dagger_{w,pq}=\sum_{\beta\gamma}[w]_{\beta\gamma}\,d^\dagger_{p\beta}d^\dagger_{q\gamma},
\label{eq:piv}
\end{equation}
where $p$ is the pivot and $q$, $j$ run over its neighbors.

The operator $B^\dagger_{w,pq}$ creates a pair of carriers on the bond $pq$ with flavor amplitude $[w]_{\beta\gamma}$ for $\beta$ on $p$ and $\gamma$ on $q$. The pivot term moves the carrier at $q$ to another neighbor $j$ of $p$ and preserves the flavor state $w$ of the pair. This is the pivot hopping term of the main text. The flavor states of a pair fall into six channels $\eta=(S,T,|T_z|)$, labeled by the total spin $S$, the total valley isospin $T$, and $|T_z|$. Channel $\eta$ contains $(2S+1)(1+|T_z|)$ orthonormal states $w$, sixteen in total, and we write $w_\eta$ for any one of them. All states of a channel have the same pivot amplitude $g_\eta$ (Table~\ref{tab:pivot}), called $g_w$ in the main text. For example, channel $(0,1,0)$, the only one with $g_\eta<0$, has a single state, the valley-symmetric intervalley spin singlet $w_{(0,1,0)}=\tfrac12\bigl(|K{\uparrow},K'{\downarrow}\rangle-|K{\downarrow},K'{\uparrow}\rangle+|K'{\uparrow},K{\downarrow}\rangle-|K'{\downarrow},K{\uparrow}\rangle\bigr)$. The first entry of each ket is the flavor on $p$, and the second that on the other site.

\begin{table}[htb]
\centering
\renewcommand{\arraystretch}{1.35}
\begin{tabular}{lc}
\hline\hline
$\eta=(S,T,|T_z|)$ & $g_\eta$ ($t^2/J_A$) \\
\hline
$(0,0,0)$ & $+\tfrac1{12}$ \\
$(0,1,0)$ & $-\tfrac{17}{12}$ \\
$(0,1,1)$ & $+\tfrac34$ \\
$(1,0,0)$ & $+\tfrac{11}{12}$ \\
$(1,1,0)$ & $+\tfrac5{12}$ \\
$(1,1,1)$ & $+\tfrac14$ \\
\hline\hline
\end{tabular}
\caption{Pivot amplitude $g_\eta$ of the six pair channels.}
\label{tab:pivot}
\end{table}

We now derive Eq.~(\ref{eq:heff}). All states with $N_c$ carriers on $N$ sites are degenerate under $H_J$, with energy $E_0=-\tfrac34J_A(N-N_c)$. We treat $H_t$ in degenerate perturbation theory within this manifold.

At first order, $H_t$ can only move an electron from a neighboring singlet, the donor, onto a carrier. A hop between two carriers would empty a site, and a hop between two singlets would triply occupy one. The constraint $n_i\in\{1,2\}$ excludes both. For both sites to remain in the manifold, the electron must carry the partner flavor $\bar\beta$ of the carrier's flavor $\beta$. The carrier then moves to the donor site with its flavor unchanged. Removing one flavor from $|s\rangle$ gives a factor $1/\sqrt2$, and re-forming $|s\rangle$ at the destination site gives another. The matrix element thus has magnitude $|t|/2$. With the fermion sign included, it is the hopping $t/2$ of Eq.~(\ref{eq:heff}) for all four flavors.

At second order the matrix element between two states $u$, $u'$ of the manifold is
\begin{equation}
H^{(2)}_{u'u}=\sum_n\frac{\langle u'|H_t|n\rangle\langle n|H_t|u\rangle}{E_0-E_n},
\label{eq:secondorder}
\end{equation}
where $n$ runs over the eigenstates of $H_J$ outside the manifold. The first hop again moves an electron from the donor onto a carrier. The donor becomes singly occupied, with energy $0$, so breaking its singlet costs $\tfrac34J_A$. The carrier site becomes a doublon other than $|s\rangle$, so the intermediate state costs $E_n-E_0=\tfrac34J_A$ for an intravalley doublon and $J_A$ for a triplet. The first hop has amplitude $|t|/\sqrt2$, again from removing one flavor from $|s\rangle$. Three flavors can arrive, since the carrier already holds flavor $\beta$. The opposite spin in the same valley forms an intravalley doublon, the same spin in the other valley forms $|t_\pm\rangle$, and the partner $\bar\beta$ forms $(|t_0\rangle\pm|s\rangle)/\sqrt2$, of which only $|t_0\rangle$ is an intermediate state. The second hop must take an electron off this doublon onto a singly occupied site. It can go back to the donor (return process) or onto a second carrier next to the first (pivot hop).

In the return process the system returns to its initial state, so the two hops enter as $|\langle n|H_t|u\rangle|^2$ and the contribution is diagonal. Each neighboring singlet contributes $\delta E=-\tfrac{17}{12}\,t^2/J_A$, the sum over the three intermediate states in Table~\ref{tab:return}. An isolated carrier has six neighboring singlets, so its energy shifts by $-\tfrac{17}{2}\,t^2/J_A$ for every flavor. This one-body term, $-\tfrac{17}{2}(t^2/J_A)\sum_in^d_i$, only shifts the chemical potential, and we omit it from Eq.~(\ref{eq:heff}). Two neighboring carriers cannot serve as donors for each other, so together they have $10$ carrier--singlet bonds instead of $12$. They lose the return energy of two singlets, which gives the repulsion $V_{\rm NN}=-2\,\delta E=+\tfrac{17}{6}\,t^2/J_A$. It is independent of the flavors of the two carriers.

\begin{table}[htb]
\centering
\renewcommand{\arraystretch}{1.35}
\begin{tabular}{lccc}
\hline\hline
$|n\rangle$ & $|\langle n|H_t|u\rangle|^2$ ($t^2$) & $1/(E_0-E_n)$ & product ($t^2/J_A$) \\
\hline
intravalley doublon & $\tfrac12$ & $-\tfrac4{3J_A}$ & $-\tfrac23$ \\
$|t_\pm\rangle$ & $\tfrac12$ & $-\tfrac1{J_A}$ & $-\tfrac12$ \\
$|t_0\rangle$ & $\tfrac14$ & $-\tfrac1{J_A}$ & $-\tfrac14$ \\[2pt]
\hline
sum & & & $-\tfrac{17}{12}$ \\
\hline\hline
\end{tabular}
\caption{Return process for one neighboring singlet: weights, denominators, and contributions.}
\label{tab:return}
\end{table}

In the pivot hop an electron moves from the donor $j$ onto the carrier at $p$. An electron then leaves $p$ for a second carrier at $q$, which must end up in $|s\rangle$. The pair of carriers thereby moves from $\{p,q\}$ to $\{p,j\}$, with $p$ acting as the pivot. The numerators have the same magnitudes as in the return process (Table~\ref{tab:return}). The Hamiltonian is SU(2) symmetric in spin and U(1)$\rtimes\mathbb Z_2$ symmetric in valley. Accordingly, $H^{(2)}$ does not mix the six channels $\eta$ and has the same value for every state within a channel. When $u$ and $u'$ are the same channel state on the pairs $\{p,q\}$ and $\{p,j\}$, this value is the pivot amplitude $g_\eta\equiv H^{(2)}_{u'u}\big|_\eta$. The operator $g_\eta\sum_{w\in\eta}B^\dagger_{w,pj}B^{\phantom\dagger}_{w,pq}$ has exactly these matrix elements, which gives Eq.~(\ref{eq:piv}). Of the six values in Table~\ref{tab:pivot}, only $g_{(0,1,0)}=-\tfrac{17}{12}\,t^2/J_A$ is negative. This channel, spin singlet $\otimes$ valley triplet with $T_z=0$, has the same flavor structure as the background singlet $|s\rangle$.

\section{Two-carrier bound state}
\label{sm:twocarrier}

In this section we solve $\hat H_{\rm eff}$ of Eq.~(\ref{eq:heff}) exactly for two carriers. In relative coordinates and at zero total momentum, this becomes the problem of a single particle hopping on the triangular lattice (see Fig.~\ref{fig:oneparticle}). The particle experiences an interaction on the six sites around the origin, and the hard-core constraint excludes the origin itself. A bound pair is a state of this particle localized near the origin. We find that two carriers bind only in channel $(0,1,0)$, as an intervalley spin-singlet pair with extended-$s$ symmetry, and only for $J_A<J_c=\tfrac{34}{3}\,|t|$. We then compute the binding energy $E_{\rm b}$, the effective mass $m^*$, and the size $R_{\rm pair}$ of this pair. Finally, we show that the pair survives a nearest-neighbor repulsion up to a critical value $V_c$.

\subsection{\texorpdfstring{Relative-coordinate equation}{Relative-coordinate equation}}

Since $\hat H_{\rm eff}$ does not mix the six channels, we fix one channel $\eta$. We write $|\mathbf r_1,\mathbf r_2\rangle$ for the state with the two carriers on sites $\mathbf r_1$ and $\mathbf r_2$ in the flavor state $w_\eta$. Exchanging the two sites gives $|\mathbf r_2,\mathbf r_1\rangle=\Pi\,|\mathbf r_1,\mathbf r_2\rangle$, with $\Pi=+1$ when $w_\eta$ is antisymmetric in the two flavors and $\Pi=-1$ when it is symmetric. A carrier is hard-core, so $\mathbf r_1\neq\mathbf r_2$. At zero total momentum the amplitude of $|\mathbf r_1,\mathbf r_2\rangle$ does not change under a common lattice translation of both carriers. It is therefore a function $\varphi(\boldsymbol\delta)$ of the relative vector $\boldsymbol\delta=\mathbf r_1-\mathbf r_2\neq\mathbf 0$, and the two-carrier state reads
\begin{equation}
|\Psi\rangle=\frac{1}{\sqrt{2N}}\sum_{\mathbf r_1\neq\mathbf r_2}\varphi(\mathbf r_1-\mathbf r_2)\,|\mathbf r_1,\mathbf r_2\rangle,\qquad \varphi(-\boldsymbol\delta)=\Pi\,\varphi(\boldsymbol\delta),
\label{eq:Psi}
\end{equation}
with $N$ the number of sites.

Taking the inner product of $\hat H_{\rm eff}|\Psi\rangle=\varepsilon|\Psi\rangle$ with $\langle\mathbf r_1,\mathbf r_2|$ gives
\begin{equation}
\varepsilon\,\varphi(\boldsymbol\delta)=t\sum_{\mathbf e}\varphi(\boldsymbol\delta+\mathbf e)+\sum_{\mathbf e'}\big[\mathsf M_\eta\big]_{\boldsymbol\delta\mathbf e'}\,\varphi(\mathbf e'),\qquad \boldsymbol\delta\neq\mathbf 0,\quad \varphi(\mathbf 0)\equiv0,
\label{eq:rel}
\end{equation}
where $\varepsilon$ is the energy of the pair and $\mathbf e$, $\mathbf e'$ run over the six nearest-neighbor vectors $\pm\mathbf e_\ell$ ($\ell=1,2,3$), of length $a$. Each carrier hops with amplitude $t/2$ in Eq.~(\ref{eq:heff}), and either hop changes $\boldsymbol\delta$ by a nearest-neighbor vector. The relative hopping amplitude is $t$, since the two contributions are equal at zero total momentum. The condition $\varphi(\mathbf 0)\equiv0$ excludes hopping onto the origin. The $V_{\rm NN}$ term and $\hat H_{\rm piv}$ act only on nearest-neighbor carriers, whose relative vector lies on the shell of six sites around the origin. On the shell, $\mathsf M_\eta$ is the $6\times6$ matrix
\begin{equation}
\mathsf M_\eta=V_{\rm NN}\,\openone_6+2g_\eta\big(\mathsf J_6-\openone_6\big),
\label{eq:shell}
\end{equation}
with $\openone_6$ the identity and $\mathsf J_6$ the matrix of all ones. The off-diagonal elements $2g_\eta$ arise from the pivot term, with either carrier as the pivot.

\begin{figure}[t]
\centering
\includegraphics{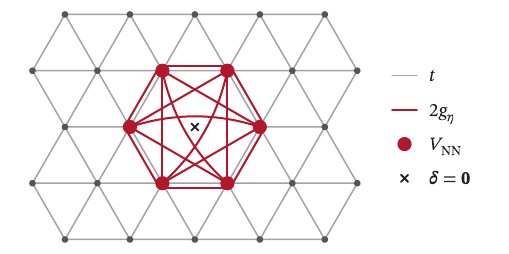}
\caption{\label{fig:oneparticle}Single-particle hopping problem of Eq.~(\ref{eq:H1p}), to which the two-carrier problem is mapped at zero total momentum. Each site corresponds to a relative vector $\boldsymbol\delta$ of the two carriers, and the particle hops between nearest-neighbor sites with amplitude $t$ (gray lines). The hard-core constraint removes the origin $\boldsymbol\delta=\mathbf 0$ (cross). On the six sites around the origin (red circles), the particle has the potential energy $V_{\rm NN}$, and the pivot term gives an additional hopping $2g_\eta$ between any two of them (red lines).}
\end{figure}

We write Eq.~(\ref{eq:rel}) as $(\hat H_{\rm hc}+\hat V_\eta)|\varphi\rangle=\varepsilon|\varphi\rangle$, with $\hat H_{\rm hc}$ the hopping term, in which the hard-core constraint excludes the origin, and $\hat V_\eta$ the interaction. Here ${|\varphi\rangle=\sum_{\boldsymbol\delta\neq\mathbf 0}\varphi(\boldsymbol\delta)|\boldsymbol\delta\rangle}$, where $|\boldsymbol\delta\rangle$ is the state with relative displacement $\boldsymbol\delta$.
The two operators read
\begin{equation}
\hat H_{\rm hc}=t\sum_{\langle\boldsymbol\delta\boldsymbol\delta'\rangle}\bigl(|\boldsymbol\delta\rangle\langle\boldsymbol\delta'|+{\rm H.c.}\bigr),\qquad
\hat V_\eta=V_{\rm NN}\sum_{\mathbf e}|\mathbf e\rangle\langle\mathbf e|+2g_\eta\sum_{\mathbf e\neq\mathbf e'}|\mathbf e\rangle\langle\mathbf e'|,
\label{eq:H1p}
\end{equation}
where $\langle\boldsymbol\delta\boldsymbol\delta'\rangle$ runs over the nearest-neighbor pairs of sites other than the origin.
On the shell, the particle has the potential energy $V_{\rm NN}$ and an additional hopping $2g_\eta$ between any two of the six sites, as illustrated in Fig.~\ref{fig:oneparticle}.

\subsection{\texorpdfstring{Interaction on the shell}{Interaction on the shell}}

The matrix $\mathsf M_\eta$ has the eigenvalue $V_{\rm NN}+10g_\eta$ on the uniform mode $|\varphi_{\rm u}\rangle=6^{-1/2}\sum_{\mathbf e}|\mathbf e\rangle$ and $\lambda_\perp=V_{\rm NN}-2g_\eta$ on the five modes orthogonal to it. The uniform mode has equal weight on the six shell sites and zero weight at the origin, so it has extended-$s$ symmetry. By Eq.~(\ref{eq:Psi}), only modes with $\varphi(-\boldsymbol\delta)=\Pi\,\varphi(\boldsymbol\delta)$ are allowed. The uniform mode is even, so it is allowed only in the channels with $\Pi=+1$, whose flavor amplitude $w_\eta$ is antisymmetric. Table~\ref{tab:shell} lists the eigenvalues of the allowed modes in each channel, with $V_{\rm NN}=\tfrac{17}{6}\,t^2/J_A$ and $g_\eta$ from Table~\ref{tab:pivot}.

Only the uniform mode of channel $(0,1,0)$ has a negative eigenvalue, $\lambda_0=V_{\rm NN}+10g_{(0,1,0)}=-\tfrac{34}{3}\,t^2/J_A$. In this mode the pivot term lowers the energy by $10|g_{(0,1,0)}|=\tfrac{85}{6}\,t^2/J_A$, five times the repulsion $V_{\rm NN}$, leaving the net attraction $|\lambda_0|$.

\begin{table}[htb]
\centering
\renewcommand{\arraystretch}{1.35}
\begin{tabular}{lc@{\hspace{1.5em}}l}
\hline\hline
$\eta=(S,T,|T_z|)$ & $\Pi$ & eigenvalues of $\mathsf M_\eta$ ($t^2/J_A$, degeneracy) \\
\hline
$(0,0,0)$ & $-1$ & $+\tfrac83$ (3) \\
$(0,1,0)$ & $+1$ & $-\tfrac{34}{3}$ (1), $+\tfrac{17}{3}$ (2) \\
$(0,1,1)$ & $+1$ & $+\tfrac{31}{3}$ (1), $+\tfrac43$ (2) \\
$(1,0,0)$ & $+1$ & $+12$ (1), $+1$ (2) \\
$(1,1,0)$ & $-1$ & $+2$ (3) \\
$(1,1,1)$ & $-1$ & $+\tfrac73$ (3) \\
\hline\hline
\end{tabular}
\caption{Exchange sign $\Pi$ of the pair wave function and the eigenvalues of the modes of $\mathsf M_\eta$ allowed by $\Pi$.}
\label{tab:shell}
\end{table}

\subsection{\texorpdfstring{Binding threshold}{Binding threshold}}

Let $\hat H_0$ be the same hopping on the full lattice, including the origin. Its eigenstates are plane waves with energy
\begin{equation}
\varepsilon_{\mathbf k}=t\,\gamma_a(\mathbf k),\qquad \gamma_a(\mathbf k)=2\sum_{\ell=1}^{3}\cos(\mathbf k\cdot\mathbf e_\ell).
\label{eq:band}
\end{equation}
Since $t<0$, $\varepsilon_{\mathbf k}\ge6t=-6|t|$. The band bottom $-6|t|$, at $\mathbf k=0$, is the energy of two carriers each at the bottom of its own band $\tfrac t2\gamma_a(\mathbf k)$. Removing the origin and adding $\hat V_\eta$ on the shell do not change the continuum of Eq.~(\ref{eq:band}), whose lower edge remains at $-6|t|$. A bound state is an eigenstate of $\hat H_{\rm hc}+\hat V_\eta$ with energy below $-6|t|$.

Since $\hat H_{\rm hc}\ge-6|t|$, a channel in which $\hat V_\eta$ has no negative eigenvalue has no eigenvalue below $-6|t|$. Hence the channels other than $(0,1,0)$ have no bound state. In channel $(0,1,0)$, $\hat H_{\rm hc}$ and $\hat V_\eta$ are invariant under the rotations and reflections that leave the origin fixed. On the shell, the only state unchanged by all of them is $|\varphi_{\rm u}\rangle$. States of any other symmetry either have no shell component or have shell components on which $\hat V_\eta=\lambda_\perp>0$, so they have no bound state. For the fully symmetric (extended-$s$) states the problem becomes
\begin{equation}
\big(\hat H_{\rm hc}+\lambda_0\,|\varphi_{\rm u}\rangle\langle\varphi_{\rm u}|\big)|\varphi\rangle=\varepsilon|\varphi\rangle.
\label{eq:Hu}
\end{equation}

For $\varepsilon<-6|t|$ the Green's function $\hat G(\varepsilon)=(\varepsilon-\hat H_{\rm hc})^{-1}$ exists. Solving Eq.~(\ref{eq:Hu}) for $|\varphi\rangle$ gives $|\varphi\rangle=\lambda_0\langle\varphi_{\rm u}|\varphi\rangle\,\hat G(\varepsilon)|\varphi_{\rm u}\rangle$. Projecting onto $\langle\varphi_{\rm u}|$ yields the bound-state condition
\begin{equation}
1=\lambda_0\,G_{\rm u}(\varepsilon),\qquad G_{\rm u}(\varepsilon)\equiv\langle\varphi_{\rm u}|\hat G(\varepsilon)|\varphi_{\rm u}\rangle.
\label{eq:bound}
\end{equation}
Since $|G_{\rm u}(\varepsilon)|$ grows from $0$ as $\varepsilon$ rises from $-\infty$ toward $-6|t|$, a root exists when $|\lambda_0|\,|G_{\rm u}(\varepsilon)|>1$ in the limit $\varepsilon\to-6|t|$.

To determine $G_{\rm u}$ at the band bottom, we relate it to the Green's function of $\hat H_0$. The origin is connected to the other sites only by the hops onto the shell, $\hat H_0|\mathbf 0\rangle=\sqrt6\,t\,|\varphi_{\rm u}\rangle$. We write $\varepsilon-\hat H_0$ as a block matrix, with $\varepsilon$ on the origin, $\varepsilon-\hat H_{\rm hc}$ on the other sites, and $-\sqrt6\,t\,|\varphi_{\rm u}\rangle$ between them. Inverting the block matrix gives the diagonal matrix element at the origin
\begin{equation}
G^0_{\mathbf 0}(\varepsilon)\equiv\langle\mathbf 0|(\varepsilon-\hat H_0)^{-1}|\mathbf 0\rangle=\frac{1}{\varepsilon-6t^2\,G_{\rm u}(\varepsilon)}.
\label{eq:Gu}
\end{equation}
In the plane-wave basis, $G^0_{\mathbf 0}(\varepsilon)=N^{-1}\sum_{\mathbf k}(\varepsilon-\varepsilon_{\mathbf k})^{-1}$. For $N\to\infty$ the density of states of the two-dimensional band is nonzero at its bottom, so $G^0_{\mathbf 0}(\varepsilon)$ diverges logarithmically as $\varepsilon\to-6|t|$. Equation~(\ref{eq:Gu}) then gives $G_{\rm u}(\varepsilon)\to-6|t|/(6t^2)=-1/|t|$.

A bound state thus exists when $|\lambda_0|>|t|$. With $|\lambda_0|=\tfrac{34}{3}\,t^2/J_A$, this condition reads
\begin{equation}
J_A<J_c=\frac{34}{3}\,|t|.
\label{eq:Jc}
\end{equation}
We find numerically that the threshold is highest at zero total momentum, so no two-carrier bound state exists for $J_A\ge J_c$.

The threshold arises from the hard-core constraint. If the origin were kept, the Green's function of the uniform mode would diverge at the band bottom like $G^0_{\mathbf 0}$. Any attraction would then bind the two carriers, as usual in two dimensions. Removing the origin keeps $G_{\rm u}$ finite, so the attraction $|\lambda_0|$ must exceed the relative hopping $|t|$. Since $|\lambda_0|\propto t^2$ grows faster than $|t|$, this condition holds once $|t|$ is large enough at fixed $J_A$.

\subsection{\texorpdfstring{Binding energy and effective mass}{Binding energy and effective mass}}

For $J_A<J_c$, the binding energy is $E_{\rm b}=-6|t|-\varepsilon_{\rm b}$, where $\varepsilon_{\rm b}<-6|t|$ is the root of Eq.~(\ref{eq:bound}). Inserting $G_{\rm u}(\varepsilon_{\rm b})=1/\lambda_0$ into Eq.~(\ref{eq:Gu}) and using $\lambda_0=-J_c|t|/J_A$ gives
\begin{equation}
-\frac{1}{G^0_{\mathbf 0}(-6|t|-E_{\rm b})}=6|t|\Big(1-\frac{J_A}{J_c}\Big)+E_{\rm b},
\label{eq:Eb}
\end{equation}
which we solve numerically for each $J_A$. The binding energy decreases as $J_A$ increases and vanishes as $J_A\to J_c$ [Fig.~\ref{fig:mf}(b)].

At small total momentum $\mathbf Q$, the energy of the bound state is $\varepsilon_{\rm b}(\mathbf Q)\simeq\varepsilon_{\rm b}+|\mathbf Q|^2/(2m^*)$, which defines the effective mass $m^*$ of the pair ($\hbar=1$). A single carrier has mass $m_d=2/(3|t|a^2)$ at the bottom of its band, $\tfrac t2\gamma_a(\mathbf k)\simeq-3|t|+|\mathbf k|^2/(2m_d)$. Two free carriers thus have mass $2m_d$, which is independent of $J_A$. At second order a single carrier gains only the one-body return energy, which shifts the chemical potential and leaves $m_d$ unchanged.

Unlike $V_{\rm NN}$, the relative hopping and the pivot term move the midpoint of the two carriers, so they depend on the total momentum $\mathbf Q$. Expanding these terms to order $|\mathbf Q|^2$ and taking the expectation value in the bound state $|\varphi\rangle$ at $\mathbf Q=\mathbf 0$ gives
\begin{equation}
\frac{2m_d}{m^*}=-\frac{\langle\hat H_{\rm hc}\rangle}{6|t|}+\frac{4|g_{(0,1,0)}|}{|t|}\,P_{\rm NN},
\label{eq:mstarsplit}
\end{equation}
where $P_{\rm NN}$ is the probability that the two carriers are nearest neighbors. The first term comes from the hopping and the second from the pivot term. Since the interaction is proportional to $1/J_A$, the Hellmann--Feynman theorem expresses both terms through $E_{\rm b}$ and $dE_{\rm b}/dJ_A$, and Eq.~(\ref{eq:mstarsplit}) becomes
\begin{equation}
\frac{m^*}{2m_d}=\frac{6|t|}{6|t|+E_{\rm b}-2J_A\,dE_{\rm b}/dJ_A}.
\label{eq:mstar}
\end{equation}
Because $dE_{\rm b}/dJ_A=-J_c|t|P_{\rm NN}/J_A^2$ is negative, $m^*<2m_d$ for all $J_A<J_c$. The ratio $m^*/(2m_d)$ increases with $J_A$ [Fig.~\ref{fig:mf}(c)].

Equations~(\ref{eq:Eb}) and (\ref{eq:mstar}) take simple forms in two limits. For strong binding ($J_A\ll J_c$), the two carriers remain nearest neighbors, $|\varphi\rangle\simeq|\varphi_{\rm u}\rangle$. Then $\varepsilon_{\rm b}\simeq\lambda_0+\langle\varphi_{\rm u}|\hat H_{\rm hc}|\varphi_{\rm u}\rangle=\lambda_0-2|t|$. With $P_{\rm NN}\simeq1$, the pivot term dominates Eq.~(\ref{eq:mstarsplit}), and we obtain
\begin{equation}
E_{\rm b}\simeq\frac{34}{3}\frac{t^2}{J_A}-4|t|,\qquad \frac{m^*}{2m_d}\simeq\frac{3J_A}{17|t|}.
\label{eq:strong}
\end{equation}
Near $J_c$, where $E_{\rm b}$ is small, inserting $G^0_{\mathbf 0}(-6|t|-E)\simeq-\sqrt3\,\ln(72|t|/E)/(12\pi|t|)$ for $E\to0$ into Eq.~(\ref{eq:Eb}) gives
\begin{equation}
E_{\rm b}\simeq72|t|\exp\Big(-\frac{2\pi}{\sqrt3}\frac{J_c}{J_c-J_A}\Big),\qquad m^*\to2m_d.
\label{eq:weak}
\end{equation}
As $J_A\to J_c$, the probability $P_{\rm NN}$ vanishes, so the pivot term no longer moves the pair.

The pair is lighter than two free carriers because the pivot hop that binds it also moves it. In a pivot hop one carrier swings around the other, so the midpoint of the pair moves. This motion gives the second term of Eq.~(\ref{eq:mstarsplit}), which lowers $m^*$. A smaller $J_A$ strengthens the pivot hop, whose amplitude is proportional to $t^2/J_A$, and makes the pair both more tightly bound and lighter. In contrast, an attraction that does not move the pair is independent of $\mathbf Q$. The on-site attraction of the attractive Hubbard model is of this kind. For such an attraction, only the hopping contributes to $2m_d/m^*$, as in the first term of Eq.~(\ref{eq:mstarsplit}). Since $\langle\hat H_{\rm hc}\rangle\ge-6|t|$, this term is at most $1$, and the pair is heavier than two free particles. In the attractive Hubbard model at strong coupling, the on-site pair moves only through virtual dissociation, and its mass grows with the attraction.

\subsection{\texorpdfstring{Pair size}{Pair size}}

\begin{figure}[t]
\centering
\includegraphics{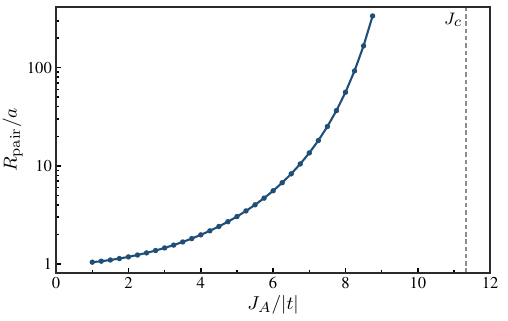}
\caption{\label{fig:twocarrier}Root-mean-square distance $R_{\rm pair}$ between the two carriers in the bound state of $\hat H_{\rm eff}$ versus $J_A$, from real-space diagonalization. The dashed line marks $J_c$.}
\end{figure}

The size of the pair is $R_{\rm pair}=\langle|\boldsymbol\delta|^2\rangle^{1/2}$, the root-mean-square distance between the two carriers in the bound state $|\varphi\rangle$. We obtain $|\varphi\rangle$ by diagonalizing Eq.~(\ref{eq:Hu}) numerically on a finite disk of sites centered at the origin, increasing the disk radius until $R_{\rm pair}$ converges. At $J_A=|t|$, $R_{\rm pair}=1.04a$, so a tightly bound pair is about one bond long (Fig.~\ref{fig:twocarrier}). The pair grows rapidly as $J_A$ approaches $J_c$, reaching $R_{\rm pair}=56a$ at $J_A=8|t|$, and $R_{\rm pair}$ diverges as $E_{\rm b}\to0$.

\subsection{\texorpdfstring{Nearest-neighbor repulsion}{Nearest-neighbor repulsion}}

To examine the effect of a nearest-neighbor repulsion between the carriers, we add the term $V\sum_{\langle ij\rangle}n^d_in^d_j$ to $\hat H_{\rm eff}$.
In the single-particle problem, $V$ raises the potential energy of the shell from $V_{\rm NN}$ to $V_{\rm NN}+V$.
The eigenvalue of the uniform mode then becomes $\lambda_0+V$, and the binding condition $|\lambda_0|>|t|$ becomes $|\lambda_0|-V>|t|$.
The bound pair therefore survives up to the critical repulsion
\begin{equation}
V_c=\frac{34}{3}\frac{t^2}{J_A}-|t|=\Bigl(\frac{J_c}{J_A}-1\Bigr)|t|.
\label{eq:Vc}
\end{equation}
We again find numerically that the threshold is highest at zero total momentum, so no two-carrier bound state exists for $V\ge V_c$.

\section{Gutzwiller approximation to the effective Hamiltonian}
\label{sm:gutzwiller}

We now turn to the case of a finite density of carriers, which we treat within a renormalized mean-field theory.
Since the carriers are hard-core fermions, we cannot treat $\hat H_{\rm eff}$ of Eq.~(\ref{eq:heff}) directly in mean-field theory.
We handle the constraint in the Gutzwiller approximation~\cite{ZhangGrosRiceShiba1988}, which multiplies each term of $\hat H_{\rm eff}$ by a factor $\kappa$ that depends on the doping $x$.
The result is a Hamiltonian $\tilde H$ of unconstrained fermions, in which the hopping and the pivot term are reduced and $V_{\rm NN}$ is unchanged.

A state that satisfies the constraint is written as $P_G|\Psi_0\rangle$, where $|\Psi_0\rangle$ is a state of unconstrained fermions.
The projector $P_G$ keeps on each site only the singlet $|s\rangle$ and the four carrier states $|\beta\rangle$.
Each term $\hat O$ of $\hat H_{\rm eff}$ changes the local configuration on the sites on which it acts from $c$ to $c'$.
In the Gutzwiller approximation the projected expectation value is the unprojected one times a factor $\kappa_O$,
\begin{equation}
\frac{\langle\Psi_0|P_G\hat OP_G|\Psi_0\rangle}{\langle\Psi_0|P_G|\Psi_0\rangle}\simeq \kappa_O\,\langle\Psi_0|\hat O|\Psi_0\rangle,\qquad \kappa_O=\sqrt{\frac{P(c)\,P(c')}{p_0(c)\,p_0(c')}}.
\label{eq:gfactor}
\end{equation}
Here $P(c)$ is the probability of the configuration $c$ in the projected state and $p_0(c)$ that in the unprojected state.
In the projected state each site is $|s\rangle$ with probability $1-x$ and a carrier with probability $x/N_f$ for each of the $N_f=4$ flavors.
Without the projection each site and flavor is occupied independently with probability $x/N_f$, and $p_0(c)$ counts only the flavors that $\hat O$ acts on.

We first consider the hopping term $\hat O=d^\dagger_{i\beta}d_{j\beta}$. In $c$, site $i$ is $|s\rangle$ and site $j$ holds a carrier of flavor $\beta$, and $c'$ is the reverse.
Then $P(c)=P(c')=(1-x)\,x/N_f$.
Since $\hat O$ acts only on flavor $\beta$ of the two sites, $p_0(c)=p_0(c')=(1-x/N_f)\,x/N_f$.
Substituting these probabilities into Eq.~(\ref{eq:gfactor}) gives
\begin{equation}
\kappa_t=\frac{1-x}{1-x/N_f}=\frac{N_f(1-x)}{N_f-x}\;\xrightarrow{N_f=4}\;\frac{4(1-x)}{4-x},
\label{eq:gt}
\end{equation}
which does not depend on $\beta$, so the entire hopping term is multiplied by the single factor $\kappa_t$.

The density term $V_{\rm NN}n^d_in^d_j$ is diagonal, $c=c'$.
The probability of two carriers on neighboring sites is $(x/N_f)^2$ with and without the projection, so its factor is $\kappa=1$ and $V_{\rm NN}$ is unchanged.

Each term $B^\dagger_{w,pj}B_{w,pq}$ of Eq.~(\ref{eq:piv}) is the product of a flavor transfer on the pivot site and a hop,
\begin{equation}
d^\dagger_{p\beta'}d^\dagger_{j\gamma'}d_{q\gamma}d_{p\beta}=F_{\beta'\beta}\,T_{\gamma'\gamma},\qquad F_{\beta'\beta}\equiv d^\dagger_{p\beta'}d_{p\beta},\quad T_{\gamma'\gamma}\equiv d^\dagger_{j\gamma'}d_{q\gamma}.
\label{eq:FT}
\end{equation}
The flavor-diagonal terms ($\beta'=\beta$), labeled D below, leave the pivot flavor unchanged. For these terms, $F_{\beta\beta}=n_{p\beta}$, the number of carriers of flavor $\beta$ on $p$. The flavor-flip terms ($\beta'\neq\beta$), labeled P, change it.
Since $P(c)$ and $p_0(c)$ factorize over sites, $\kappa$ is a product of one factor per site, $\kappa=\kappa_p\,\kappa_q\,\kappa_j$.
The sites $q$ and $j$ carry the same configurations as in the hopping term, so $\kappa_q\,\kappa_j=\kappa_t$.
On the pivot site, a flavor-diagonal term acts on the single flavor $\beta$, so $P(c)=p_0(c)=x/N_f$ and $\kappa_p=1$.
A flavor-flip term acts on both $\beta$ and $\beta'$, so $p_0(c)=(x/N_f)(1-x/N_f)$ while $P(c)=x/N_f$, and $\kappa_p=1/(1-x/N_f)$.
Hence the flavor-diagonal terms are multiplied by $\kappa_D=\kappa_t$, and the flavor-flip terms by
\begin{equation}
\kappa_P=\frac{\kappa_t}{1-x/N_f}=\frac{N_f^2(1-x)}{(N_f-x)^2}\;\xrightarrow{N_f=4}\;\frac{16(1-x)}{(4-x)^2}.
\label{eq:gP}
\end{equation}
Multiplying the D terms by $\kappa_D$ and the P terms by $\kappa_P$, and noting that their sum is the original pivot term, we obtain
\begin{equation}
P_G\,B^\dagger_{w,pj}B_{w,pq}\,P_G\simeq \kappa_P\,B^\dagger_{w,pj}B_{w,pq}-\delta\kappa\sum_\beta n_{p\beta}\sum_{\gamma\gamma'}[w]_{\beta\gamma'}[w]^*_{\beta\gamma}\,d^\dagger_{j\gamma'}d_{q\gamma},\qquad \delta\kappa\equiv \kappa_P-\kappa_D=\frac{4x(1-x)}{(4-x)^2}.
\label{eq:pivGA}
\end{equation}
The renormalized effective Hamiltonian is therefore
\begin{equation}
\tilde H=\kappa_t\,\frac t2\sum_{\langle ij\rangle\beta}\bigl(d^\dagger_{i\beta}d_{j\beta}+{\rm H.c.}\bigr)+V_{\rm NN}\sum_{\langle ij\rangle}n^d_in^d_j+\kappa_P\,\hat H_{\rm piv}-\delta\kappa\,\hat K,
\label{eq:Htilde}
\end{equation}
where $\hat K$ is the flavor-diagonal part of $\hat H_{\rm piv}$,
\begin{equation}
\hat K=\sum_\eta g_\eta\sum_{w\in\eta}\sum_p\sum_{\substack{q,j\,\sim\,p\\ j\neq q}}\sum_\beta n_{p\beta}\sum_{\gamma\gamma'}[w]_{\beta\gamma'}[w]^*_{\beta\gamma}\,d^\dagger_{j\gamma'}d_{q\gamma}.
\label{eq:K}
\end{equation}

\section{Mean-field theory of the renormalized Hamiltonian}
\label{sm:meanfield}

We treat $\tilde H$ of Eq.~(\ref{eq:Htilde}) in mean-field theory with a uniform, nonmagnetic ansatz.

The order parameters are the nearest-neighbor hopping amplitude $\chi_a$ and the nearest-neighbor pair amplitude $\Phi_{\rm NN}$ in the attractive channel $(0,1,0)$,
\begin{equation}
\begin{aligned}&\langle d^\dagger_{i\beta}d_{i\gamma}\rangle=\delta_{\beta\gamma}\,\frac x{N_f},\qquad \langle d^\dagger_{i\beta}d_{j\gamma}\rangle=\delta_{\beta\gamma}\,\chi_a,\\&\langle d_{i\beta}d_{j\gamma}\rangle=\Phi_{\rm NN}\,[w]_{\beta\gamma},\qquad w\equiv w_{(0,1,0)},\end{aligned}\qquad j\sim i.
\label{eq:ansatz}
\end{equation}
A uniform $\Phi_{\rm NN}$ gives the pairing the extended-$s$ symmetry of the two-carrier bound state.

The Gutzwiller factors impose the hard-core constraint only on average.
A paired state of the unconstrained fermions still has pairs of carriers on the same site, with the on-site pair amplitude $\Phi_0$ defined by
\begin{equation}
\langle d_{i\beta}d_{i\gamma}\rangle=\Phi_0\,[w]_{\beta\gamma}.
\label{eq:onsite}
\end{equation}
This is an artifact of the mean-field calculation, since the constraint forbids two carriers on the same site.
We remove it by requiring $\Phi_0=0$.
We impose this condition with a Lagrange multiplier $D_0$, which adds the on-site pairing term $D_0\sum_{i\beta\gamma}[w]_{\beta\gamma}\,d^\dagger_{i\beta}d^\dagger_{i\gamma}+{\rm H.c.}$ to the mean-field Hamiltonian.

We decouple the quartic terms of $\tilde H$ into Hartree, Fock, and pairing terms using Eq.~(\ref{eq:ansatz}). After the Fourier transform $d_{i\beta}=N^{-1/2}\sum_{\mathbf k}e^{i\mathbf k\cdot\mathbf r_i}d_{\mathbf k\beta}$, the mean-field Hamiltonian is, up to a constant,
\begin{equation}
\hat H_{\rm MF}=\sum_{\mathbf k\beta}\xi_{\mathbf k}\,d^\dagger_{\mathbf k\beta}d_{\mathbf k\beta}+\frac12\sum_{\mathbf k}\sum_{\beta\gamma}\Bigl(D(\mathbf k)\,[w]_{\beta\gamma}\,d^\dagger_{\mathbf k\beta}d^\dagger_{-\mathbf k\gamma}+{\rm H.c.}\Bigr).
\label{eq:HMF}
\end{equation}
The gap function is
\begin{equation}
D(\mathbf k)=2D_0-\lambda\,\Phi_{\rm NN}\,\gamma_a(\mathbf k),\qquad \gamma_a(\mathbf k)=2\sum_{\ell=1}^{3}\cos(\mathbf k\cdot\mathbf e_\ell),
\label{eq:gap}
\end{equation}
where $\lambda=V_{\rm NN}+10\,\kappa_P\,g_{(0,1,0)}$ and $\mathbf e_\ell$ are the three nearest-neighbor vectors of Eq.~(\ref{eq:band}).
The dispersion is
\begin{equation}
\xi_{\mathbf k}=\Bigl[\kappa_t\,\frac t2-\bigl(V_{\rm NN}-10\,\delta\kappa\,g_{(1,1,1)}\bigr)\chi_a\Bigr]\gamma_a(\mathbf k)+\kappa_t\,\frac{17}{12}\frac{t^2}{J_A}\,\frac x{N_f}\,\gamma_a(\mathbf k)^2-\mu.
\label{eq:xiD}
\end{equation}
In $\xi_{\mathbf k}$, the $\gamma_a(\mathbf k)$ term collects the renormalized hopping and the Fock terms.
The $\gamma_a(\mathbf k)^2$ term is the Hartree contribution from the pivot terms, a hop between two neighbors of the pivot.
The chemical potential $\mu$, fixed by $x$, absorbs the uniform on-site terms.

Diagonalizing $\hat H_{\rm MF}$ gives the quasiparticle energy
\begin{equation}
E_{\mathbf k}=\sqrt{\xi_{\mathbf k}^2+|D(\mathbf k)|^2/4}.
\label{eq:Ek}
\end{equation}
In the ground state of $\hat H_{\rm MF}$, the expectation values of Eq.~(\ref{eq:ansatz}) give the self-consistent equations
\begin{equation}
x=\frac{N_f}{2N}\sum_{\mathbf k}\Bigl(1-\frac{\xi_{\mathbf k}}{E_{\mathbf k}}\Bigr),\qquad \chi_a=\frac1{12N}\sum_{\mathbf k}\gamma_a(\mathbf k)\Bigl(1-\frac{\xi_{\mathbf k}}{E_{\mathbf k}}\Bigr),\qquad \Phi_{\rm NN}=\frac1{12N}\sum_{\mathbf k}\gamma_a(\mathbf k)\,\frac{D(\mathbf k)}{E_{\mathbf k}}.
\label{eq:selfcons}
\end{equation}
The condition $\Phi_0=0$ reads
\begin{equation}
\frac1{2N}\sum_{\mathbf k}\frac{D(\mathbf k)}{E_{\mathbf k}}=0,
\label{eq:constraint}
\end{equation}
which fixes $D_0$.
Writing $2D_0=\lambda\,\Phi_{\rm NN}\,c$, we obtain
\begin{equation}
D(\mathbf k)=-\lambda\,\Phi_{\rm NN}\,[\gamma_a(\mathbf k)-c],\qquad c=\frac{\sum_{\mathbf k}\gamma_a(\mathbf k)/E_{\mathbf k}}{\sum_{\mathbf k}1/E_{\mathbf k}}.
\label{eq:d0c}
\end{equation}

For a solution with $\Phi_{\rm NN}\neq0$, the third relation in Eq.~(\ref{eq:selfcons}) and Eq.~(\ref{eq:constraint}) give the gap equation
\begin{equation}
1=-\frac{\lambda}{12N}\sum_{\mathbf k}\frac{[\gamma_a(\mathbf k)-c]^2}{E_{\mathbf k}}.
\label{eq:gapeq}
\end{equation}
The sum is positive, so a pairing solution requires $\lambda<0$.
Inserting ${g_{(0,1,0)}=-\tfrac{17}{12}\,t^2/J_A}$ (Table~\ref{tab:pivot}) and ${V_{\rm NN}=\tfrac{17}{6}\,t^2/J_A}$ gives $\lambda=\tfrac{17}{6}\,(1-5\kappa_P)\,t^2/J_A$.
Hence $\lambda<0$ means $\kappa_P>\tfrac15$, that is, $x<4\sqrt{85}-36\approx0.88$, independent of $J_A$.

Every quantity depends on $\mathbf k$ only through $\gamma_a(\mathbf k)$, so the momentum sums become integrals over the density of states of $\gamma_a(\mathbf k)$, which we evaluate for $N\to\infty$.
We solve Eqs.~(\ref{eq:selfcons}) and (\ref{eq:constraint}) for $\chi_a$, $\Phi_{\rm NN}$, $\mu$, and $D_0$ at fixed $x$.
Figure~\ref{fig:mf}(d) shows the resulting $\Phi_{\rm NN}$.

\subsection{\texorpdfstring{Dilute limit}{Dilute limit}}

The mean-field theory reproduces the exact two-carrier solution of Sec.~\ref{sm:twocarrier} in this limit.
As $x\to0$, the Gutzwiller factors approach $1$, and $\chi_a$ and the $\gamma_a(\mathbf k)^2$ term of Eq.~(\ref{eq:xiD}) vanish.
The coupling $\lambda$ becomes the eigenvalue $\lambda_0$ of the uniform mode, and $\xi_{\mathbf k}=(\varepsilon_{\mathbf k}-2\mu)/2$ with $\varepsilon_{\mathbf k}$ of Eq.~(\ref{eq:band}).
Since $\Phi_{\rm NN}\to0$, $E_{\mathbf k}\to\xi_{\mathbf k}>0$, and Eq.~(\ref{eq:gapeq}) becomes
\begin{equation}
1=\frac{\lambda_0}{6N}\sum_{\mathbf k}\frac{[\gamma_a(\mathbf k)-c]^2}{2\mu-\varepsilon_{\mathbf k}}.
\label{eq:dilute}
\end{equation}
With $c$ of Eq.~(\ref{eq:d0c}), $\gamma_a(\mathbf k)=\varepsilon_{\mathbf k}/t$, and $\sum_{\mathbf k}\gamma_a(\mathbf k)=0$, the sum equals $N\,[2\mu-1/G^0_{\mathbf 0}(2\mu)]/t^2$, which is $6N\,G_{\rm u}(2\mu)$ by Eq.~(\ref{eq:Gu}).
Equation~(\ref{eq:dilute}) is thus the bound-state condition~(\ref{eq:bound}) of two carriers with the energy $\varepsilon=2\mu$.
The constant $c$ removes the on-site component of the pair wave function, as the hard-core constraint removes the origin in Eq.~(\ref{eq:rel}).
A solution exists only for $J_A<J_c=\tfrac{34}{3}\,|t|$, and then $2\mu=-6|t|-E_{\rm b}$ with the exact binding energy $E_{\rm b}$ of Eq.~(\ref{eq:Eb}).
Without the constraint, $c=0$ and the sum in Eq.~(\ref{eq:dilute}) diverges as $2\mu$ approaches the band bottom $-6|t|$.
A solution would then exist for any $\lambda_0<0$, even for $J_A>J_c$.

In this limit the pair amplitude $\langle d_{\mathbf k\beta}d_{-\mathbf k\gamma}\rangle=F(\mathbf k)\,[w]_{\beta\gamma}$ is $F(\mathbf k)=D(\mathbf k)/(2\xi_{\mathbf k})$.
It is proportional to the wave function of the two-carrier bound state in momentum space.
For $N_f=4$, the first relation in Eq.~(\ref{eq:selfcons}) becomes $x=N^{-1}\sum_{\mathbf k}F(\mathbf k)^2$, and the third gives $\Phi_{\rm NN}$ as the nearest-neighbor component of $F$.
All carriers are bound into pairs, and $6\Phi_{\rm NN}^2/x$ is the probability $P_{\rm NN}$ of Eq.~(\ref{eq:mstarsplit}) that the two carriers in a pair are nearest neighbors.
Hence
\begin{equation}
\Phi_{\rm NN}=\sqrt{P_{\rm NN}\,x/6},
\label{eq:PhiNNdilute}
\end{equation}
where $P_{\rm NN}$ depends only on $J_A$.
Since $P_{\rm NN}$ decreases from $1$ for a tightly bound pair to $0$ at $J_c$, $\Phi_{\rm NN}$ at small $x$ decreases as $J_A$ increases [Fig.~\ref{fig:mf}(d)].

\subsection{\texorpdfstring{Pairing threshold}{Pairing threshold}}

\begin{figure}[t]
\centering
\includegraphics{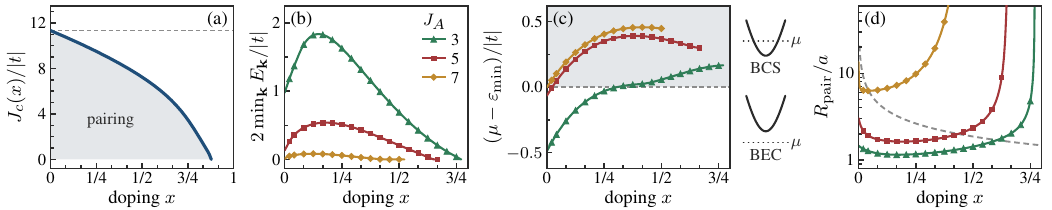}
\caption{\label{fig:mfsm}Mean-field theory with the constraint $\Phi_0=0$. (a) Pairing threshold $J_c(x)$; pairing occurs for $J_A<J_c(x)$ (shaded). The dashed line marks the two-carrier threshold $J_c=\frac{34}{3}|t|$. (b) Single-particle gap $2\min_{\mathbf k}E_{\mathbf k}$, (c) chemical potential $\mu$ measured from the bottom $\varepsilon_{\min}$ of the carrier band (shaded), and (d) size $R_{\rm pair}$ of the Cooper pair, versus doping $x$ for $J_A/|t|=3$, $5$, and $7$, up to the doping where the pairing ends. The sketches beside (c) show $\mu$ (dotted line) inside the carrier band (BCS) and below its bottom (BEC). The dashed line in (d) is the mean distance between pairs, $\ell=(\sqrt3/x)^{1/2}a$.}
\end{figure}

At finite $x$, pairing occurs only below a threshold $J_c(x)$.
Since $\xi_{\mathbf k}$ of Eq.~(\ref{eq:xiD}) depends on $\mathbf k$ only through $\gamma_a(\mathbf k)$, the Fermi surface of the normal state is a contour $\gamma_a(\mathbf k)=\gamma_F$.
As $\Phi_{\rm NN}\to0$, both sums in Eq.~(\ref{eq:d0c}) diverge logarithmically at the Fermi surface, so $c\to\gamma_F$ and the gap function vanishes on the Fermi surface.
Unlike the BCS gap equation, Eq.~(\ref{eq:gapeq}) then has no logarithmic divergence, and a solution with $\Phi_{\rm NN}\neq0$ exists only when
\begin{equation}
-\frac{\lambda}{12N}\sum_{\mathbf k}\frac{[\gamma_a(\mathbf k)-\gamma_F]^2}{|\xi_{\mathbf k}|}>1,
\label{eq:threshold}
\end{equation}
with $\xi_{\mathbf k}$ of the normal state.
This condition reads $J_A<J_c(x)$ [Fig.~\ref{fig:mfsm}(a)].
The threshold equals the two-carrier value $\tfrac{34}{3}\,|t|$ at $x\to0$ and decreases with $x$.
The Gutzwiller factor $\kappa_P$ reduces the pivot hopping, because fewer singlets are left on the neighboring sites at larger $x$, while the repulsion $V_{\rm NN}$ is unchanged.
The attraction $|\lambda|$ therefore weakens with $x$, and the threshold vanishes at $x\approx0.88$, where $\lambda$ changes sign.

For $J_A<J_c(x)$, we find a single solution with $\Phi_{\rm NN}\neq0$, and its energy $\langle\tilde H\rangle$ is lower than that of the normal state.
Its $\Phi_{\rm NN}$ follows Eq.~(\ref{eq:PhiNNdilute}) at small $x$ and shows a dome [Fig.~\ref{fig:mf}(d)].
It vanishes at the doping where $J_c(x)=J_A$ and is exponentially small close to it.
Since $J_c(x)$ decreases with $x$, the pairing ends at a smaller doping for a larger $J_A$.
For $J_A=10|t|$, close to $J_c$, the pairing is limited to small $x$ and is exponentially weak.
In contrast, DMRG finds pairing at $J_A=10|t|$ and $x=1/4$ [Fig.~\ref{fig:2}], above the mean-field threshold $J_c(1/4)=9.5|t|$.

\subsection{\texorpdfstring{Single-particle gap}{Single-particle gap}}

In mean-field theory, adding or removing a carrier costs at least $\min_{\mathbf k}E_{\mathbf k}$, so the analog of the DMRG gap $\Delta_{1e}$ is $2\min_{\mathbf k}E_{\mathbf k}$ [Fig.~\ref{fig:mfsm}(b)].
At $x\to0$ it equals the binding energy $E_{\rm b}$.
It increases with $x$ because the gap function $D(\mathbf k)$ grows with $\Phi_{\rm NN}$.
At larger $x$ it decreases and vanishes where the pairing ends.
This dome is qualitatively similar to the DMRG result of Fig.~\ref{fig:2}(b), whose maximum lies at $x\approx1/4$--$1/3$.

\subsection{\texorpdfstring{BEC--BCS crossover}{BEC--BCS crossover}}

Figure~\ref{fig:mfsm}(c) shows the chemical potential measured from the bottom $\varepsilon_{\min}=\min_{\mathbf k}(\xi_{\mathbf k}+\mu)$ of the carrier band.
At $x\to0$, $\mu-\varepsilon_{\min}=-E_{\rm b}/2$, and the carriers form a dilute gas of bound pairs.
As $x$ increases, $\mu$ rises and enters the band, as in a BCS superconductor.
A larger $J_A$ weakens the binding and moves the system toward the BCS limit, so $\mu$ enters the band at a smaller doping.

Figure~\ref{fig:mfsm}(d) shows the size $R_{\rm pair}$ of the Cooper pair, defined by $R_{\rm pair}^2=\sum_{\mathbf r}|\mathbf r|^2F(\mathbf r)^2/\sum_{\mathbf r}F(\mathbf r)^2$.
Here $F(\mathbf r)$ is the Fourier transform of the pair amplitude $F(\mathbf k)=D(\mathbf k)/(2E_{\mathbf k})$.
At $x\to0$, $R_{\rm pair}$ is the size of the two-carrier bound state (Fig.~\ref{fig:twocarrier}).
It first decreases with $x$, and then grows and diverges where the pairing ends, as it does for two carriers at $J_c$.
At small $x$ the pairs are smaller than the mean distance between them, $\ell=(\sqrt3/x)^{1/2}a$ [dashed line in Fig.~\ref{fig:mfsm}(d)].
As $x$ increases, $\ell$ shrinks and the pairs overlap.
This happens at a smaller doping for a larger $J_A$, where the pairs are larger.
Doping thus drives a crossover from a Bose--Einstein condensate (BEC) of tightly bound pairs to a BCS-like state of overlapping pairs.
The mean-field solution for $J_A/|t|=3$ and $5$ lies in the crossover regime at $x\approx1/4$--$1/3$, where the DMRG gap is largest [Fig.~\ref{fig:2}(b)].
Note that the superconductor in TBG was also found to be close to the BEC--BCS crossover~\cite{Cao2018SC}.

\subsection{\texorpdfstring{Gap structure and tunneling spectra}{Gap structure and tunneling spectra}}

\begin{figure}[t]
\centering
\includegraphics{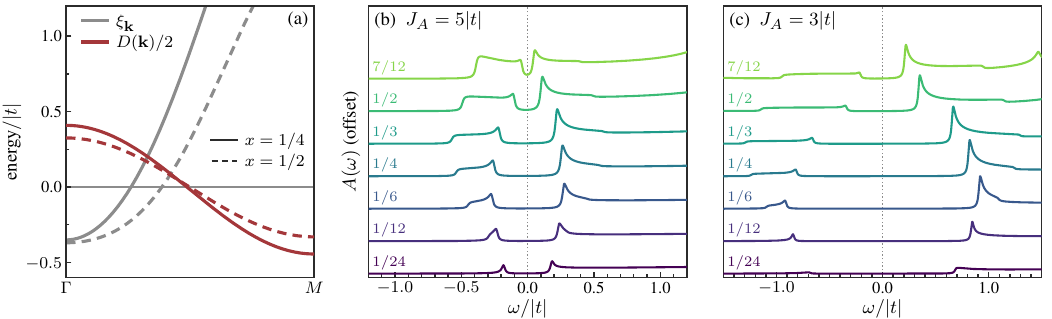}
\caption{\label{fig:mfdos}(a) Dispersion $\xi_{\mathbf k}$ and gap function $D(\mathbf k)/2$ along $\Gamma$--$M$ for $J_A=5|t|$ at $x=1/4$ (solid) and $1/2$ (dashed). (b),(c) Carrier density of states $A(\omega)$ for (b) $J_A=5|t|$ and (c) $J_A=3|t|$ at $x=1/24$, $1/12$, $1/6$, $1/4$, $1/3$, $1/2$, and $7/12$ (bottom to top), offset vertically.}
\end{figure}

The gap function of Eq.~(\ref{eq:d0c}) changes sign on the contour $\gamma_a(\mathbf k)=c$ [Fig.~\ref{fig:mfdos}(a)].
The constraint~(\ref{eq:constraint}) requires this sign change.
This contour and the Fermi surface are both contours of $\gamma_a(\mathbf k)$, so they do not cross, and $|D(\mathbf k)|$ is constant on the Fermi surface.
We find $c<\gamma_F$ in the paired state, so the sign change lies outside the Fermi surface.
There $\xi_{\mathbf k}\neq0$, so $E_{\mathbf k}$ remains finite even where $D(\mathbf k)$ vanishes.
The quasiparticle spectrum is therefore fully gapped, without nodes.
The two contours meet only at the threshold, where $c\to\gamma_F$.

Figures~\ref{fig:mfdos}(b) and \ref{fig:mfdos}(c) show the carrier density of states
\begin{equation}
A(\omega)=\frac{N_f}{N}\sum_{\mathbf k}\bigl[u_{\mathbf k}^2\,\delta(\omega-E_{\mathbf k})+v_{\mathbf k}^2\,\delta(\omega+E_{\mathbf k})\bigr],\qquad u_{\mathbf k}^2,\,v_{\mathbf k}^2=\frac12\Bigl(1\pm\frac{\xi_{\mathbf k}}{E_{\mathbf k}}\Bigr),
\label{eq:dos}
\end{equation}
with a Lorentzian broadening of half-width $0.012|t|$.
Since a carrier is a doped hole, the electron tunneling spectrum is the mirror image $A(-\omega)$, up to overall factors.
The spectra are U-shaped, with a flat gap and sharp coherence peaks.
The gap first widens and then narrows with $x$, following the dome of Fig.~\ref{fig:mfsm}(b).
At small $x$, most of the weight lies in the carrier-addition peak at $\omega>0$, as expected for bound pairs.
The removal peak at $\omega<0$ grows with $x$ as the state becomes BCS-like.
The step at negative $\omega$ is the bottom of the carrier band.
These fully gapped spectra contrast with the V-shaped spectra observed in TBG~\cite{Oh2021STM} and twisted trilayer graphene~\cite{Park2026Tunnel}.
A nodal gap may require the two orbitals of TBG in addition to spin and valley~\cite{ZhaoZhang2025}.

\section{Momentum distribution at large $J_A$}
\label{sm:nk}

To characterize the normal state at large $J_A$, we measure the momentum distribution $n(\mathbf k)$ at $x=1/4$ for $J_A=50$, $100$, and $200$ with iDMRG on the $L_y=3$ cylinder [Fig.~\ref{fig:2}(e)].
If the Fermi surface encloses only the carriers, as in the sFL of the main text, we expect a small hole pocket around $\Gamma$, whose size counts the $x$ carriers per site.
A conventional Fermi liquid would instead have a large Fermi surface.

The three sites around the circumference of the cylinder form a ring.
We label the momentum by $k_x$, the phase between neighboring rings, and $k_y$, the phase between neighboring sites of a ring.
Since a ring has three sites, $k_y$ takes only the values $0$ and $\pm2\pi/3$, so $n(\mathbf k)$ splits into three branches labeled by $k_y$.
The distribution reads
\begin{equation}
n(\mathbf k)=\sum_{\mathbf r}e^{-i\mathbf k\cdot\mathbf r}\sum_\alpha\langle\tilde c^\dagger_{\mathbf 0\alpha}\tilde c_{\mathbf r\alpha}\rangle,
\label{eq:nk}
\end{equation}
where the site $\mathbf r$ lies $i_x$ rings and $i_y$ sites away from the site $\mathbf 0$, so that $\mathbf k\cdot\mathbf r=i_xk_x+i_yk_y$.
In Cartesian components, $\mathbf r=i_x\mathbf e_1+i_y\mathbf e_2$ with $\mathbf e_1=(\tfrac{\sqrt3}{2},\tfrac12)$ and $\mathbf e_2=(0,1)$, and $k_x\equiv\mathbf k\cdot\mathbf e_1$, $k_y\equiv\mathbf k\cdot\mathbf e_2$.

The iDMRG setup is the same as that used in Fig.~\ref{fig:2}(c), with an eight-ring unit cell and $\chi=4000$.
In Eq.~(\ref{eq:nk}), $\mathbf 0$ is averaged over the sites of the first ring of the unit cell, and the sum runs over $|i_x|\le24$.
Because $\tilde c_{i\alpha}$ annihilates only electrons on doubly occupied sites, the $\mathbf r=\mathbf 0$ term equals $2(1-x)$ rather than the density $2-x$, so the $n(\mathbf k)$ defined here differs from the electron momentum distribution by the constant $-x$, which does not affect the pocket.

In the $k_y=0$ branch, $n(\mathbf k)$ shows a hole pocket: it is depleted only around $\Gamma$ [blue curves in Fig.~\ref{fig:2}(e)].
The pocket edge, identified by the maximum of $|\partial n/\partial k_x|$, lies at $k_F/\pi\approx0.19$ for all three $J_A$.
The $k_y=\pm2\pi/3$ branches, mirror images of each other under $k_x\to-k_x$, are featureless [gray curves in Fig.~\ref{fig:2}(e)].

The pocket is formed by the carriers, whose band $\tfrac t2\gamma_a(\mathbf k)$ from Eqs.~(\ref{eq:heff}) and (\ref{eq:band}) is lowest at $\Gamma$.
Its size is set by the carriers alone: with $x$ carriers per site, a ring holds $3x/4$ carriers per flavor, which fill the $k_y=0$ branch up to $k_F/\pi=3x/4\approx0.19$, as observed.
In contrast, a conventional Fermi liquid counts all $2-x$ electrons, more than one per flavor per ring, which a single branch cannot hold.
The remaining electrons would partially fill the $k_y=\pm2\pi/3$ branches, giving each branch an edge, which the gray curves in Fig.~\ref{fig:2}(e) do not show.

\begin{figure}[t]
\centering
\includegraphics{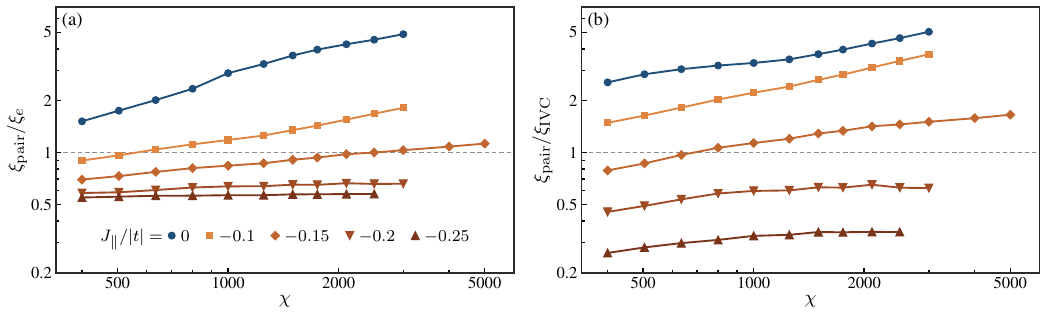}
\caption{\label{fig:jpar}Ratio of the pair correlation length $\xi_{\rm pair}$ to (a) the single-particle correlation length $\xi_e$ and (b) the intervalley coherent (IVC) correlation length $\xi_{\rm IVC}$ versus $\chi$, at $x=1/4$ on $L_y=3$ cylinders, from iDMRG with a sixteen-ring unit cell, for $J_A=1$ and $J_\parallel=0$, $-0.1$, $-0.15$, $-0.2$, and $-0.25$. The dashed lines mark a ratio of one.}
\end{figure}

\section{Intersite flavor exchange $J_\parallel$}
\label{sm:jpar}

To examine the effect of the intersite flavor exchange $J_\parallel$ on the pairing, we add to Eq.~(\ref{eq:H}) the term
\begin{equation}
H_\parallel=J_\parallel\sum_{\langle ij\rangle}\Bigl(\frac12\sum_{\alpha\alpha'}c^\dagger_{i\alpha}c_{i\alpha'}\,c^\dagger_{j\alpha'}c_{j\alpha}-\frac18\,n_in_j\Bigr).
\label{eq:hpar}
\end{equation}
For $J_\parallel<0$, it lowers the ten flavor-symmetric states of two neighboring singlons below the six flavor-antisymmetric ones by $|J_\parallel|$.
The extended-$s$ pair is flavor antisymmetric, so a negative $J_\parallel$ works against the pairing.
With iDMRG on $L_y=3$ cylinders at $x=1/4$ and $J_A=1$, we measure the pair and single-particle correlation lengths $\xi_{\rm pair}$ and $\xi_e$ for $J_\parallel=0$, $-0.1$, $-0.15$, $-0.2$, and $-0.25$.
We also measure the correlation length $\xi_{\rm IVC}$ of the intervalley coherent (IVC) operator $c^\dagger_{iK\uparrow}c_{iK'\uparrow}$.
These three correlation lengths follow from the leading transfer-matrix eigenvalues in the sectors $(N_K,N_{K'},2S^z)=(1,1,0)$, $(1,0,1)$, and $(1,-1,0)$, respectively.

For $|J_\parallel|\le0.15$, both ratios $\xi_{\rm pair}/\xi_e$ and $\xi_{\rm pair}/\xi_{\rm IVC}$ increase with $\chi$, as at $J_\parallel=0$ [Fig.~\ref{fig:jpar}].
For $|J_\parallel|\ge0.2$, both level off below one.
At $x=1/6$ and $1/3$ as well, both ratios increase with $\chi$ at $|J_\parallel|=0.15$ and level off below one at $|J_\parallel|=0.2$.
These results indicate that the pairing nevertheless survives a small $|J_\parallel|$ and is lost only for $|J_\parallel|\ge0.2$.

\end{document}